\documentclass[twocolumn,secnumarabic,amssymb, nobibnotes, aps, prl, superscriptaddress]{revtex4-2}

\usepackage[]{graphicx}
\usepackage{tabularx}
\usepackage[usenames,dvipsnames]{color}
\usepackage{soul}
\usepackage{ulem}
\usepackage{bm}
\usepackage{amsmath}
\usepackage{lineno}
\usepackage{gensymb}

\usepackage{times}
\usepackage{amsfonts}
\usepackage{mathrsfs}
\usepackage{graphicx}% Include figure files
\usepackage{dcolumn}% Align table columns on decimal point
\usepackage{bm}% bold math
\usepackage{color}

\usepackage[colorlinks,bookmarks=false,citecolor=blue,linkcolor=red,urlcolor=blue]{hyperref}
\usepackage{multirow}
\usepackage{physics}
\usepackage{ulem}

\begin{document}

%\title{Nonreciprocal supercurrent by vortex phase texture}

\title{Supercurrent diode with high winding vortex}

\author{Yuri Fukaya}

\email{fukaya.yuri@spin.cnr.it}
\affiliation{CNR-SPIN, I-84084 Fisciano (SA), Italy, c/o Universit\'a di Salerno, I-84084 Fisciano (SA), Italy}
\affiliation{Faculty of Environmental Life, Natural Science and Technology, Okayama University, 700-8530 Okayama, Japan}

\author{Maria Teresa Mercaldo}
\affiliation{Dipartimento di Fisica ``E. R. Caianiello", Universit\`a di Salerno, IT-84084 Fisciano (SA), Italy}

\author{Daniel Margineda}
%\email{daniel.margineda@nano.cnr.it}
\affiliation{NEST Istituto Nanoscienze-CNR and Scuola Normale Superiore, I-56127, Pisa, Italy}

\author{Alessandro Crippa}
\affiliation{NEST Istituto Nanoscienze-CNR and Scuola Normale Superiore, I-56127, Pisa, Italy}

\author{Elia Strambini}
\affiliation{NEST Istituto Nanoscienze-CNR and Scuola Normale Superiore, I-56127, Pisa, Italy}

\author{Francesco Giazotto}
%\email{francesco.giazotto@sns.it}
\affiliation{NEST Istituto Nanoscienze-CNR and Scuola Normale Superiore, I-56127, Pisa, Italy}

\author{Carmine Ortix}
\affiliation{Dipartimento di Fisica ``E. R. Caianiello", Universit\`a di Salerno, IT-84084 Fisciano (SA), Italy}

\author{Mario Cuoco}
%\email{mario.cuoco@spin.cnr.it}
\affiliation{CNR-SPIN, I-84084 Fisciano (SA), Italy, c/o Universit\'a di Salerno, I-84084 Fisciano (SA), Italy}

\begin{abstract}
Nonreciprocal supercurrent refers to the phenomenon where the maximum dissipationless current in a superconductor depends on its direction of flow. This asymmetry underlies the operation of superconducting diodes and is often associated with the presence of vortices. Here, we investigate supercurrent nonreciprocal effects in a superconducting weak-link hosting distinct types of vortices. 
    We demonstrate how the winding number of the vortex, its spatial configuration, and the shape of the superconducting lead can steer the sign and amplitude of the supercurrent rectification. 
We identify a general criterion for optimizing the rectification amplitude based on vortex patterns, focusing on configurations where the first harmonic of the supercurrent vanishes.
    We prove that supercurrent nonreciprocal effects can be used to diagnose high-winding vortex and to distinguish between different types of vorticity.
    Our results provide a toolkit
    for controlling supercurrent rectification through vortex phase textures and detecting unconventional vortex states.
\end{abstract}

\maketitle

\noindent\large{\textbf{Introduction}}\normalsize\\ 
Vortices represent fundamental topological excitations in superfluids and superconductors. They have been predicted and successfully observed in a broad range of systems, including superconductors \cite{Abrikosov:1956sx}, liquid helium \cite{Vinen1961,Bewley2008,Gomez2014}, ultracold atomic gases \cite{Hadzibabic2006}, photon fields \cite{Allen1992}, and exciton-polariton condensates \cite{Lagoudakis2008,Roumpos2011}.
Vortices are generally characterized by a quantized phase winding and a suppressed order parameter at their core.
A gradient of the phase $\phi$ of the superconducting order parameter $\Delta=|\Delta| \exp (i \phi)$ yields a circulating supercurrent around the core of the vortex, whereas the amplitude is vanishing, i.e.\ $|\Delta|\rightarrow 0$. For conventional superconductors, due to the single-valuedness of the superconducting order parameter, the winding number associated with the phase gradient is forced to be an integer $V_0$. In principle, $V_0$ can assume values 
larger than one. 
Apart from having $V_0$ vortices with winding number equal to one, a giant vortex with winding number $V_0$ can also be realized.  
A giant vortex is expected to be relevant in small superconductors with confined geometries. To date, probing of such unconventional vortex state has been mostly addressed by magnetic means in suitably tailored geometric configurations, e.g.\ by Hall probe \cite{Geim1997,Geim2000,Morelle2004} and scanning SQUID microscopy \cite{Kokubo2010}, or by tunneling microscopy and spectroscopy \cite{Kanda2004,Cren2009,Cren2011,Timmermans2016,Samokhvalov2019}.
\\
Here, we unveil a specific relation between the occurrence of vortex states with any given winding number in a Josephson weak-link and the rectification of the supercurrent flowing across the junction. 
Supercurrent rectification is a timely problem at the center of intense investigation \cite{and20,baur22, bau22, wu22,jeo22,nad23,Ghosh2024}. A large body of work devoted to supercurrent rectification focuses on superconducting states marked by linear phase gradients with, for instance, Cooper pair momentum \cite{lin22,pal22,yuan22}, spin-flipper by ferromagnets~\cite{Pal_2019}, or helical phases \cite{Edelstein95,Ilic22,dai22,he22,Turini22}, as well as screening currents \cite{hou23,sun23}, and supercurrent related to self-field \cite{kras97,GolodNatComms2022} or back-action mechanisms~\cite{margineda2023backaction}. 
Vortices or circular phase gradients, associated to conventional winding, are also expected to yield supercurrent diode effects, due to the induced Josephson phase shift \cite{Golod2010}, and their role has been investigated for a variety of physical configurations \cite{GolodNatComms2022,sur22,GutfreundNatComms2023,Gillijns07,Ji21,He19,MarginedaCommunPhys2023,Paolucci23,Greco23,Lustikova2018,Itahashi20}. 
However, whether and how superconducting phase patterns with nontrivial winding for the vorticity can be probed by nonreciprocal response, which are problems not yet fully uncovered. 
\\
By studying nonreciprocal supercurrent effects arising from distinct types of vortex phase texture in Josephson weak-links, in this paper, we show that while a superconducting phase with vortices can generally lead to nonvanishing supercurrent rectification, the sign and amplitude of the rectification in the Josephson diode effect can be manipulated by the position of the vortex core and the winding of the phase vortex. 
We thereby uncover a general criterion to single out which phase vortex configuration can maximize the rectification amplitude of the supercurrent. Our findings provide a toolkit for the design and control of supercurrent rectification by vortex phase texture.

\noindent\large{\textbf{Model and methodology}}\normalsize\\
In this section, we present the model Hamiltonian in real space for the superconducting leads, along with the methodology we used to analyze the current-phase relationship and to identify the maximum supercurrent that can flow in both directions across the junction.

\begin{figure}[t!]
    \centering
    \includegraphics[width=8.6cm]{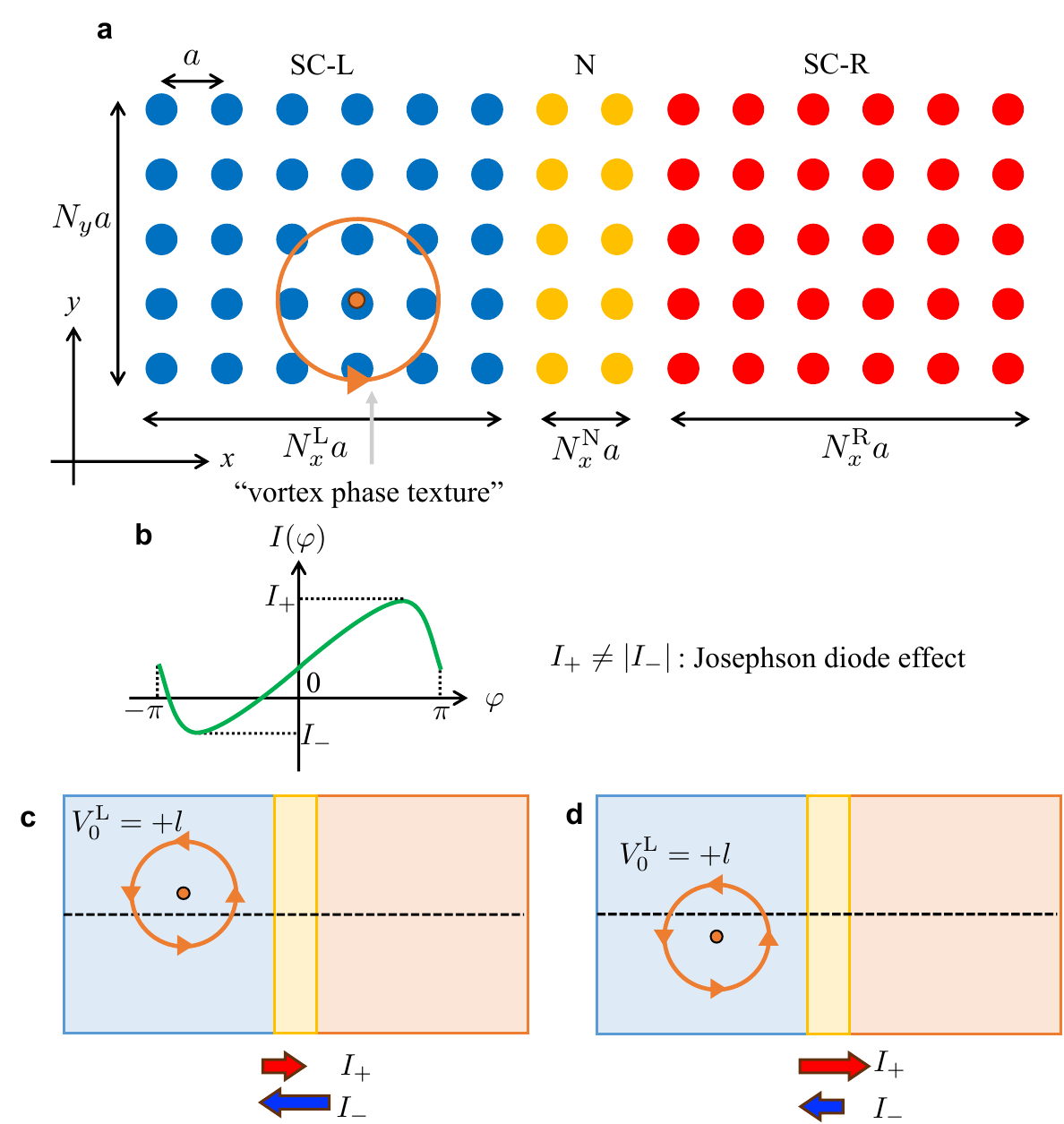}
    \caption{\textbf{Image of Josephson junctions with a vortex.}
    (a) Schematic of a Josephson junction in the lattice model with a vortex phase texture.
    SC-L, N, and SC-R mean the Left-side superconductor (SC), the Normal metal, and the Right-side SC.
    $N^\mathrm{L}_{x}$, $N^\mathrm{R}_{x}$, $N^\mathrm{N}_{x}$, $N_{y}$, and $a$ denote the number of sites in the left and right-side SCs, and in the normal metal along the $x$-direction, and number of sites along the $y$-direction, and lattice constant.
    $\bm{r}^\mathrm{L}_{0}=(x^\mathrm{L}_{0},y^\mathrm{L}_{0})$ is the core position.
    (b) A representative nonreciprocal current phase relation with $\varphi$ being the phase bias across the junction.
    (c)(d) Sketch of the superconducting weak link with a vortex placed on the left side of the junction. 
    We show two representative vortex positions along the lateral direction.
    Since the supercurrent pattern is spatially modified by the phase vortex, it can become nonreciprocal, i.e.\ the forward supercurrent $I_+$ is different from the backward one $I_{-}$. The vortex winding, $V^\mathrm{L}_{0}$, can take any integer number $l$.}
    \label{fig:1}
\end{figure}%

\noindent
\textbf{Model Hamiltonian.}
We consider a planar superconducting weak-link with the geometry shown in Fig.~\ref{fig:1} (a). For convenience, the system size is expressed as $N_{y}a\times(N^\mathrm{L}_{x}+N^\mathrm{N}_{x}+N^\mathrm{R}_{x})a$ with $N_{y}$ being the lateral number of sites, $N^\mathrm{L,R,N}_{x}$ the site numbers in the left ($L$), right ($R$) and normal ($N$) region of the superconducting weak-link, and $a$ is the lattice length, respectively. The superconducting order parameter on the left and right side of the junction is given by $\Delta_\mathrm{L,R}e^{i\varphi_\mathrm{L,R}}$.

The Hamiltonian describing the superconducting junction illustrated in Fig.~\ref{fig:1} (a) is written as
\begin{align}
    \hat{\mathcal{H}}=\sum_{ j_{x},j_{y}}\sum_{j'_{x},j'_{y}}\hat{C}^{\dagger}(j_{x},j_{y})\hat{H}(j_{x},j_{y};j'_{x},j'_{y};\varphi)\hat{C}(j'_{x},j'_{y}),
\end{align}%
with $\varphi=\varphi_\mathrm{L}-\varphi_\mathrm{R}$ being the phase difference between the superconducting order parameters in the two sides of the junction
and $\bm{j}=(j_{x},j_{y})$ with $j_{x}\in[-N^\mathrm{L}_{x}+1,N^\mathrm{N}_{x}+N^\mathrm{R}_{x}]$ and $j_{y}\in[-(N_y-1)/2,(N_y-1)/2]$ indicating the site indices in the real space.
Here, $\langle j_{x},j_{y};j'_{x},j'_{y}\rangle$ indicates the summation within the nearest-neighbor hopping and $\hat{C}^{\dagger}(j_{x},j_{y})=[c^{\dagger}_{j_{x},j_{y},\uparrow},c_{j_{x},j_{y},\downarrow}]$ denotes the creation operator at $\bm{j}$.
Because we do not consider any symmetry breaking in the normal state, the number of the basis can be $2(N^\mathrm{L}_{x}+N^\mathrm{N}_{x}+L^\mathrm{R}_{x})N_{y}$, not $4(N^\mathrm{L}_{x}+N^\mathrm{N}_{x}+L^\mathrm{R}_{x})N_{y}$.
Then $\hat{H}(j_{x},j_{y};j'_{x},j'_{y};\varphi)$ is given by
\begin{align}
    \hat{H}(j_{x},j_{y};j'_{x},j'_{y};\varphi)&=\hat{H}^\mathrm{L}+\hat{H}^\mathrm{R}+\hat{H}^\mathrm{N}+\hat{H}^\mathrm{L}_{J}+\hat{H}^\mathrm{R}_{J},
    \label{full_H}
\end{align}%
with the Hamiltonian in the Nambu space $\hat{H}^\mathrm{L,N,R}$ and the tunneling part $\hat{H}^\mathrm{L,R}_{J}$. 
The detail of the Hamiltonian is described by
\begin{align}
    \hat{H}^\mathrm{L}&=\sum^{0}_{j_{x}=-N^\mathrm{N}_{x}+1}\sum^{(N_{y}-1)/2}_{j_{y}=-(N_{y}-1)/2}[(-\varepsilon)c^{\dagger}_{j_{x},j_{y},\uparrow}c_{j_{x},j_{y},\uparrow}\notag\\
    &-tc^{\dagger}_{j_{x},j_{y}+1,\uparrow}c_{j_{x},j_{y},\uparrow}-tc^{\dagger}_{j_{x}+1,j_{y},\uparrow}c_{j_{x},j_{y},\uparrow}\notag\\
    &+\Delta_\mathrm{L}(j_{x},j_{y})c_{j_{x},j_{y},\uparrow}c_{j_{x},j_{y},\downarrow}]+h.c.,
\end{align}%
\begin{align}
    \hat{H}^\mathrm{N}&=\sum^{N^\mathrm{N}_{x}}_{j_{x}=1}\sum^{(N_{y}-1)/2}_{j_{y}=-(N_{y}-1)/2}[(-\varepsilon)c^{\dagger}_{j_{x},j_{y},\uparrow}c_{j_{x},j_{y},\uparrow}\notag\\
    &-tc^{\dagger}_{j_{x},j_{y}+1,\uparrow}c_{j_{x},j_{y},\uparrow}-tc^{\dagger}_{j_{x}+1,j_{y},\uparrow}c_{j_{x},j_{y},\uparrow}]+h.c.,
\end{align}%
\begin{align}
    \hat{H}^\mathrm{R}&=\sum^{N^\mathrm{N}_{x}+N^\mathrm{R}_{x}}_{j_{x}=N^\mathrm{N}_{x}+1}\sum^{(N_{y}-1)/2}_{j_{y}=-(N_{y}-1)/2}[(-\varepsilon)c^{\dagger}_{j_{x},j_{y},\uparrow}c_{j_{x},j_{y},\uparrow}\notag\\
    &-tc^{\dagger}_{j_{x},j_{y}+1,\uparrow}c_{j_{x},j_{y},\uparrow}-tc^{\dagger}_{j_{x}+1,j_{y},\uparrow}c_{j_{x},j_{y},\uparrow}\notag\\
    &+\Delta_\mathrm{R}(j_{x},j_{y};\varphi)c_{j_{x},j_{y},\uparrow}c_{j_{x},j_{y},\downarrow}]+h.c.,
\end{align}%
with the on-site energy term $\varepsilon=-0.25t$, the energy gap function with spin-singlet $s$-wave state $\Delta_\mathrm{L,R}e^{i\varphi_\mathrm{L,R}}$, and the nearest neighbor hopping terms $t_{x,y}$.
In the present study, the pairing amplitude is on-site and included in the local term of the Hamiltonian for the left and right-side superconductors.
We point out that the position dependence of the order parameter $\Delta_\mathrm{R}(j_{x},j_{y};\varphi)$ takes into account the change in the phase and amplitude due to the presence of the vortices. This is a convenient approach that can be generally applied for analyzing the effects of nonstandard vortices without explicitly incorporating the source responsible for generating the vortex.
The tunneling Hamiltonian is expressed as
\begin{align}
    \hat{H}^\mathrm{L}_{J}=\sum^{(N_{y}-1)/2}_{j_{y}=-(N_{y}-1)/2}[-t_\mathrm{int}tc^{\dagger}_{1,j_{y},\uparrow}c_{0,j_{y},\uparrow}]+h.c.,
\end{align}%
\begin{align}
    \hat{H}^\mathrm{R}_{J}&=\sum^{(N_{y}-1)/2}_{j_{y}=-(N_{y}-1)/2}[-t_\mathrm{int}tc^{\dagger}_{N^\mathrm{N}_{x}+1,j_{y},\uparrow}c_{N^\mathrm{N}_{x},j_{y},\uparrow}]\notag\\
    &+h.c.,
\end{align}%
with $t_\mathrm{int}=0.90$ the charge transfer amplitude at the interface, setting out the transparency of the junction.

\begin{figure*}[t!]
    \centering
    \includegraphics[width=17cm]{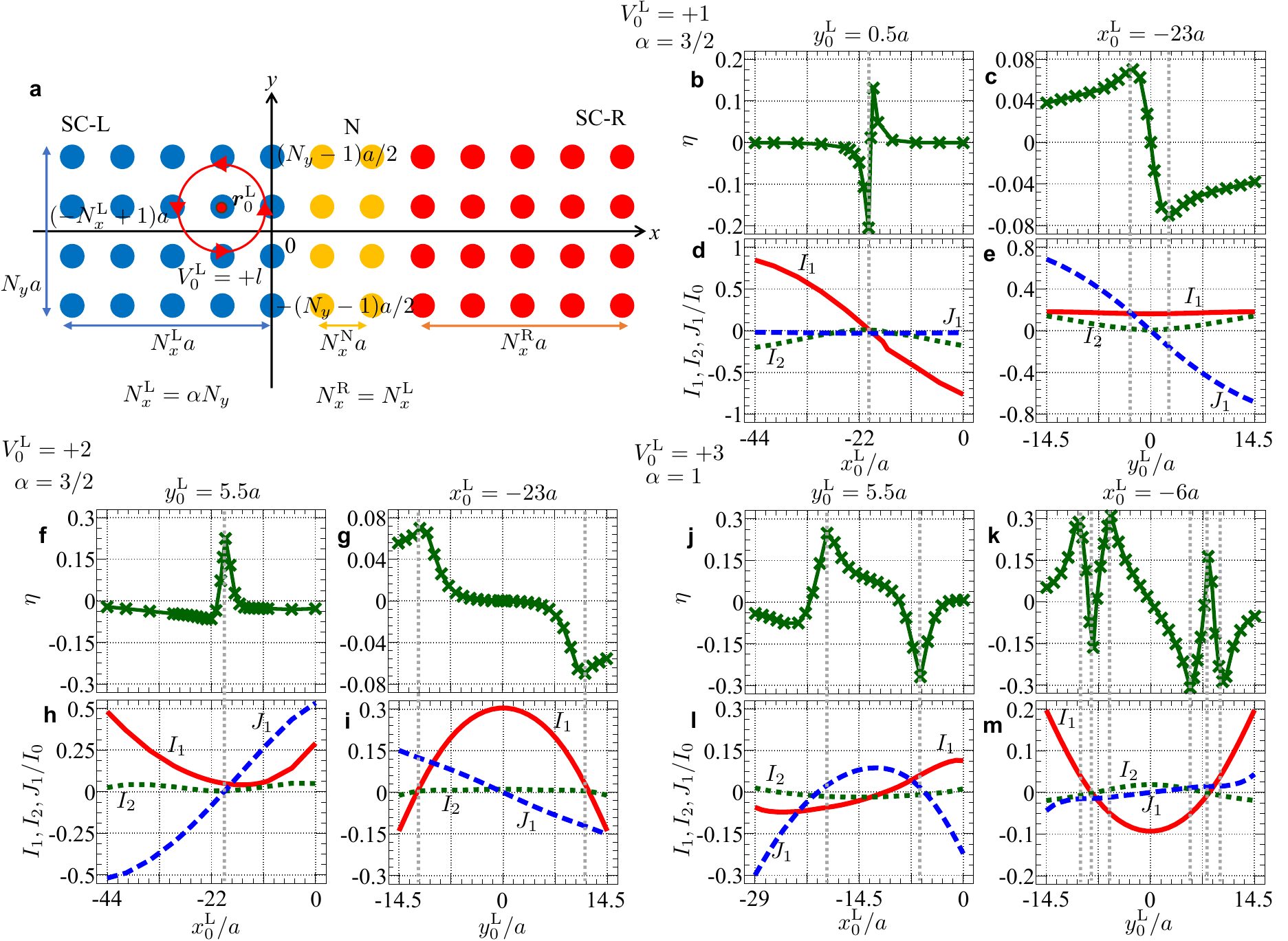}
    \caption{\textbf{Space dependence of rectification amplitudes with a vortex.}
    (a) Schematic illustration of the Josephson junction with real space coordinates. 
    The analysis is performed for a vortex configuration having winding number $V^\mathrm{L}_{0}=+l$ with $l=1,2,3$.
    SC-L, N, and SC-R mean the Left-side superconductor (SC), the Normal metal, and the Right-side SC.
    $N^\mathrm{L}_{x}$, $N^\mathrm{R}_{x}$, $N^\mathrm{N}_{x}$, $N_{y}$, and $a$ denote the number of sites in the left and right-side SCs, and in the normal metal along the $x$-direction, and number of sites along the $y$-direction, and lattice constant.
    Evolution of the rectification amplitude $\eta$ with regard to the vortex core coordinates for different winding numbers: (b,c) $V^\mathrm{L}_0=1$, (f,g) $V^\mathrm{L}_0=2$, and (j,k) $V^\mathrm{L}_0=3$. 
    Spatially resolved harmonics of the supercurrent: $I_{1}$ (red-solid line), $I_{2}$ (green-dotted), and $J_{1}$ (blue-dashed) indicate the odd-parity first harmonic, the odd-parity second harmonic, and the even-parity first harmonic amplitude, respectively. 
    Longitudinal scan: $I_{1}$, $I_{2}$, and $J_{1}$ at a given $y^\mathrm{L}_{0}$ as a function of $x^\mathrm{L}_{0}$ for (d) $V^\mathrm{L}_0=1$, (h) $V^\mathrm{L}_0=2$, and (l) $V^\mathrm{L}_0=3$. 
    Lateral scan: $I_{1}$, $I_{2}$, and $J_{1}$ at a given $x^\mathrm{L}_{0}$ as a function of $y^\mathrm{L}_{0}$ for (e) $V^\mathrm{L}_0=1$, (i) $V^\mathrm{L}_0=2$, and (m) $V^\mathrm{L}_0=3$.
    The gray dotted lines refer to the position of the maximal rectification and are a guide to indicate the values $I_{1}$, $I_{2}$, and $J_{1}$. We set $y^\mathrm{L}_{0}$ as (b)(d) $0.5a$ and (f)(h)(j)(l) $5.5a$, and $x^\mathrm{L}_{0}$ as (c)(e)(g)(i) $-23a$ and (k)(m) $-6a$.
    The aspect ratio is (b)-(i) $\alpha=3/2$ and (j)-(m) $\alpha=1$.
    The maximal rectification $\eta$ occurs for vortex core positions corresponding to a supercurrent with $I_{1}$, $I_{2}$, and $J_{1}$ components that are comparable in size. The sign change of $\eta$ is related to the vanishing of $J_{1}$ and to the zeros of $I_{1}$ when the amplitude is comparable to $J_{1}$. Multiple sign reversals of $\eta$ are observed for $V^\mathrm{L}_0=3$.
    Parameters: $|\Delta_0|=0.02t$ (superconducting energy gap amplitude), $t_\mathrm{int}=0.90$ (transparency at the interface), $N^\mathrm{N}_{x}=10$, $N_{y}=30$, and $z^\mathrm{L}_{0}=10a$ (vortex size).}
    \label{fig:2}
\end{figure*}

\noindent
\textbf{Josephson current and rectification.}
Next, we consider how the Josephson current flowing in the junction [Fig.~\ref{fig:1} (a)] is evaluated.
The current phase relation of the Josephson current is obtained by evaluating the variation of the free energy $F$ with respect to the phase bias across the junction, i.e.\ $\varphi$:
\begin{align}
    I(\varphi)=2\times\frac{2e}{\hbar}\frac{\partial F(\varphi)}{\partial \varphi},
\end{align}%
with the free energy in a Josephson junction at zero temperature, evaluated as
\begin{align}
    F(\varphi)&=\frac{1}{N_{x} N_{y}}\sum_{E<0}E(\varphi),
\end{align}%
\begin{align}
    \hat{H}(j_{x},j_{y};j'_{x},j'_{y};\varphi)|\Phi\rangle&=E(\varphi)|\Phi\rangle.    
\end{align}%
Here, $N_{x}=N^\mathrm{L}_{x}+N^\mathrm{N}_{x}+N^\mathrm{R}_{x}$ and $N_{y}$ denote the total number of sites along the $x$-direction and along the $y$-direction, and $E(\varphi)$ and $|\Phi\rangle$ stand for the eigenvalue and eigenstate of $\hat{H}(x,y;x',y';\varphi)$.
We numerically obtain $E(\varphi)$ by the full diagonalization of $\hat{H}(j_{x},j_{y};j'_{x},j'_{y};\varphi)$ in Eq.\ (\ref{full_H}). 
The computational analysis is performed at zero temperature, but a thermal change does not qualitatively alter the results.
For our purposes, we recall that if time-reversal is broken, then the supercurrent can be expanded in even and odd harmonics with respect to the phase bias variable as 
\begin{align}
    I(\varphi)=\sum_{m}[I_{m}\sin (m\varphi)+J_{m}\cos (m\varphi)],
\end{align}%
in Ref.~\cite{GolubovRMP2004}.
To assess the nonreciprocal supercurrent due to the presence of the vortex in the junction, we evaluate the rectification amplitude $\eta$ that is conventionally expressed as
\begin{align}
    \eta=\frac{I_{+}-|I_{-}|}{I_{+}+|I_{-}|},
\end{align}%
with $I_{+(-)}$ the maximum amplitude of the supercurrent for forward (backward) directions, respectively.
For the presented results, we set the energy gap amplitude as $|\Delta_{0}|=0.02t$ ($t$ is the electron hopping amplitude), the maximum Josephson current value, without vortex, as $I_\mathrm{0}=0.012|\Delta_{0}|(\frac{2e}{\hbar})$, and the size of the superconducting leads as  $N^\mathrm{L}_{x}=N^\mathrm{R}_{x}=\alpha N_{y}$ with $\alpha$ being the aspect ratio setting out a square or rectangular shape of the superconductors in the junction. 
The computation is performed for $N_{y}=30$, and $z^\mathrm{L}_{0}=10a$ for the size of the vortex core.
A variation of these lengths does not affect the results; thus, for clarity, we introduce the coordinates $(j_{x},j_{y})$ in the Josephson junction [Fig.~\ref{fig:2} (a)].

\begin{figure*}[t!]
    \centering
    \includegraphics[width=13.5cm]{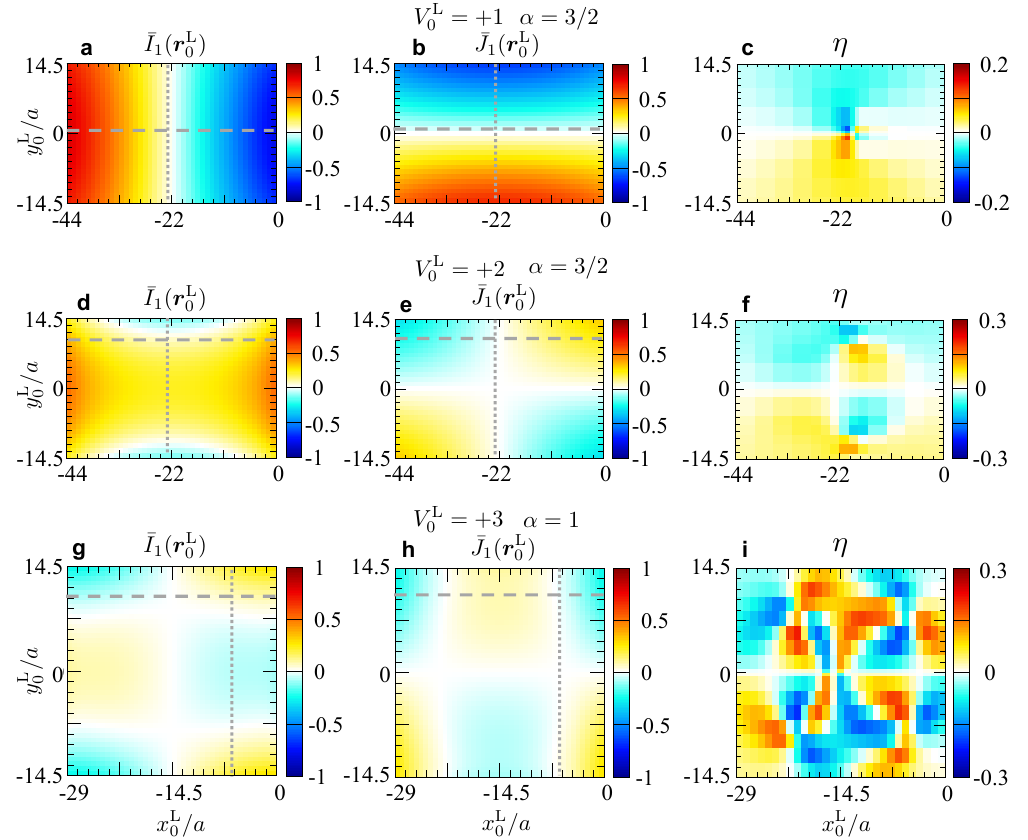}
    \caption{\textbf{Evaluation of the first and second harmonics and the rectifications as a function of the space with a vortex.}
    Amplitude of the first harmonics of the Josephson current and the rectification versus the vortex core positions $\bm{r}^\mathrm{L}_{0}=(x^\mathrm{L}_{0},y^\mathrm{L}_{0})$ for different vortex winding $V^\mathrm{L}_0$: (a,b,c) $V^\mathrm{L}_0=1$, (d,e,f) $V^\mathrm{L}_0=2$, (g,h,i) $V^\mathrm{L}_0=3$. 
    $\bar{I}_{1}$, $\bar{J}_{1}$, and $\eta$ indicate the odd- and even-parity first harmonics of the supercurrent evaluated by the direct Cooper pairs tunneling and the amplitude of the rectification.
    The color bars indicate the amplitude of (a,d,g) $\bar{I}_1(\bm{r}^\mathrm{L}_{0})$, (b,e,h) $\bar{J}_1(\bm{r}^\mathrm{L}_{0})$, and (c,f,i) the rectification amplitude $\eta$.
    In (a,b,d,e,g,h), gray-dotted and dashed lines stand for the vortex core positions,  $y^\mathrm{L}_{0}$ and $x^\mathrm{L}_{0}$, evaluated in Fig.~\ref{fig:2}.
    The aspect ratio is (a)-(f) $\alpha=3/2$ and (g)-(i) $\alpha=1$.
    In (c)(f)(i), the rectification amplitude is evaluated by scanning all the positions of the vortex cores by performing the computation of the supercurrent for the weak link assuming the following parameters: $|\Delta_0|=0.02t$ (superconducting energy gap amplitude), $t_\mathrm{int}=0.90$ (transparency at the interface), $N^\mathrm{N}_{x}=10$, $N_{y}=30$, and $z^\mathrm{L}_{0}=10a$ (vortex size).}
\label{fig:3}
\end{figure*}

\noindent\large{\textbf{Supercurrent Rectification
%superconducting lead larger than the vortex phase coherence
}}\normalsize\\ 
In this section, we study the supercurrent rectification by examining one and two vortices situated within the superconducting lead of the junction and assuming that the core position can be varied and it can be marked by different winding numbers. 
We would like to point out that to realize nonreciprocal superconducting phenomena, it is essential to break both inversion and time-reversal symmetries. This condition is crucial for establishing nonreciprocal behavior in superconducting transport, and it can be directly inferred in a superconducting weak link, for example, from the parity properties of the current-phase relation with respect to the phase bias.
In the examined system, these symmetries are broken by the presence of vortices in the superconducting leads. Specifically, when analyzing the junction with either one or two vortices, the system lacks a center of inversion and is not invariant under time-reversal symmetry.
For example, considering a single vortex on one side of the junction, there are no inversion centers, and the phase pattern does not remain invariant under time reversal. These symmetry considerations directly influence the current-phase relation of the Josephson junction. Specifically, the relation features a first harmonic with even parity, which implicitly reflects the broken time-reversal symmetry. Additionally, the second harmonic contains both even- and odd-parity components with respect to the applied phase bias. Although the odd-parity component can exist regardless of the vortex presence when the junction is outside the tunneling regime, the even-parity contribution arises from the vortex-induced symmetry breaking. Then, we note that the rectification amplitude is generally nonzero when a nontrivial even-parity first harmonic coexists with the second harmonic component. Our results thus will demonstrate that the vortex plays a role in the rectification process of the supercurrent by generating a nonzero even-parity first harmonic term. In particular, the position of the vortex affects both the amplitude of the first harmonic term and the second harmonic component, too, thus resulting in a modulation of the sign and amplitude of the supercurrent rectification.
Here, we adopt a superconducting phase profile designed to qualitatively capture the main features of the giant vortex as observed in relevant experiments \cite{Fink1968,Bruyndoncx1999,Kanda2004,Kramer_2009_giant,Cren2011} and discussed in several theoretical studies \cite{Buzdin1993_quanta,Schweigert1998,Kaori_Tanaka_2002,ChaoPRB2009,Palonen_2013,Amundsen2016,Liu2021}. However, this approach does not incorporate the boundary-induced corrections required to fully satisfy supercurrent conservation at the sample edges. While such corrections are known to play a crucial role in stabilizing high-winding-number vortices, particularly in finite-size systems \cite{Schweigert1998,Kaori_Tanaka_2002,ChaoPRB2009,Palonen_2013,Amundsen2016,Liu2021} or in the presence of strong pinning \cite{Buzdin1993_quanta}, we point out that a self-consistent treatment of the superconducting phase, accounting for boundary effects, lies beyond the scope of the present study.

\noindent
\textbf{Single vortex configuration.}
To investigate the rectification properties of the supercurrent in the junction, we start by considering a single vortex configuration assuming a variable winding number and core position as illustrated in Fig.~\ref{fig:1}.
The vortex is placed on the left side of the junction, but the results for vortices on the right side of the junction can be directly obtained by applying an inversion symmetry transformation and considering that it leads to a sign variation of the rectification amplitude (see Supplementary Info - Section B). 
Antivortex configurations through mirror and time-reversal transformations can also be directly deduced from the results of the single vortex (see Supplementary Info - Section B).
To simulate the vortex state with the core at a given position, $\bm{r}^\mathrm{L}_{0}=(x_{0}^\mathrm{L},y_{0}^\mathrm{L})$, the phase value at the site $\bm{j}$ in the left-side superconductor is given by (Fig. ~\ref{fig:2} (a))
\begin{align}
    \varphi^\mathrm{L}_\mathrm{v}(\bm{j},\bm{r}^\mathrm{L}_{0})=V^\mathrm{L}_{0}\arg[(j_{x}a-x_{0}^\mathrm{L})+i(j_{y}a-y_{0}^\mathrm{L})], \label{phiv}
\end{align}%
with $V^\mathrm{L}_{0}$ being the winding number of the vortex.
$\varphi^\mathrm{L}_\mathrm{v}(\bm{j},\bm{r}^\mathrm{L}_{0})$ reverses its sign by changing the sign of $V^\mathrm{L}_{0}$, and $|V^\mathrm{L}_{0}|$ corresponds to the number of sign changes in the real space $\bm{j}$ for the phase value when winding around the core of the vortex.
The pair potential $\Delta_\mathrm{L}$ and $\Delta_\mathrm{R}$ for spin-singlet $s$-wave state are given by
\begin{align}
    \Delta_\mathrm{L}=|\Delta_{0}|
    \tilde\Theta(\bm{j},\bm{r}^\mathrm{L}_{0})    e^{i\varphi^\mathrm{L}_\mathrm{v}(\bm{j},\bm{r}^\mathrm{L}_{0})}  \label{DeltaL}
\end{align}%
and $\Delta_\mathrm{R}=|\Delta_{0}|$ 
with $|\Delta_{0}|=0.02t$ being the superconducting energy gap amplitude.
In the presence of a phase vortex texture, the amplitude of the pair potential $\Delta_\mathrm{L}$ is modified by $\tilde\Theta^\mathrm{L}(\bm{j},\bm{r}^\mathrm{L}_{0})$ for each site $\bm{j}$:
\begin{align}
    \tilde\Theta^\mathrm{L}(\bm{j},\bm{r}^\mathrm{L}_{0})=\tanh\left[\frac{|\bm{j}a-\bm{r}^\mathrm{L}_{0}|}{z^\mathrm{L}_0}\right],
\end{align}%
with $z^\mathrm{L}_{0}=10a$ being the vortex size.
\\
In Figs.~\ref{fig:2} (b)-(m), we display the rectification amplitude for various vortex winding values while altering the position of the vortex core. We consider $\alpha=3/2$ and $\alpha=1$ as representative cases for the aspect ratio of the superconducting lead.
The analysis for all vortex winding is performed by scanning the vortex core position within the superconducting lead along the longitudinal ($x$) and lateral direction $y$.
We start by considering a conventional vortex with winding $V^\mathrm{L}_{0}=1$. 
The outcome of the study indicates that the rectification tends to vanish and changes the sign if the vortex core is placed at the crossing of the longitudinal and transverse mirror lines of the left superconducting lead [Figs.~\ref{fig:2} (b,c)]. For the geometry of the junction, these symmetry lines correspond to the $y\sim 0$ and $x\sim -22a$ axes. Other vortex core positions away from the symmetry lines give a negligible rectification. One can also observe that the maximum of the rectification occurs nearby these nodal points at a distance that is set by the vortex core size $z^\mathrm{L}_{0}$.
The behavior of the supercurrent rectification can be understood by inspection of the amplitude of the first harmonics in the current phase relation. For instance, placing the vortex core along the longitudinal direction at a different distance from the interface, we find that the value of the first odd-parity harmonic ($I_1$) changes sign at $x\sim -22a$ [Figs.~\ref{fig:2} (d)]. Then, this configuration allows for a sign change of the time-conserving component of the supercurrent and thus to an effective $0$-$\pi$ Josephson phase transition. For such a position, one can observe that the even-parity first harmonic ($J_1$) and the odd-parity second harmonic ($I_2$) components have a small amplitude and are comparable to that of $I_1$. 
Instead, when considering the evolution of $\eta$ as a function of the lateral coordinate $y$ we find that the sign change of $\eta$ occurs nearby $y\sim 0$. 
Now, the sign reversal is guided by the first even-parity harmonic term ($J_1$) in the current phase relation [Figs.~\ref{fig:2} (e)]. 
The $J_1$ component sets out the amplitude of the spontaneous supercurrent induced by the presence of the vortex at zero applied phase bias (i.e.\ $\varphi=0$). Both the scan along the $x$ and $y$ directions indicate that the rectification amplitude is maximal (of the order of 20$\%$) for vortex core positions $\bm{r}^\mathrm{L}_{0}$ that correspond to configurations for which the first harmonics of the supercurrent have comparable strength. 
This tendency towards optimal rectification of the supercurrent is an expected outcome that can be directly inferred from the analysis of a Josephson system exhibiting current-phase relationships with generic yet comparable amplitudes for ($I_1$), ($I_2$), and ($J_1$) (\cite{ytanakaPRB2022}) (see Supplementary Info - Section B). In Section C of the Supplementary Info, we present the current-phase relationship profiles for several representative vortex configurations. The harmonic content depicted in Fig. 2 has been derived from the current-phase relationships evaluated for each corresponding vortex configuration. 
\\
Moving to even high-winding, $V^\mathrm{L}_0=2$, we find that the maximal amplitude of the rectification occurs for vortex core positions that are now closer to the lateral edges of the superconductor but still midway from the interface [Figs.~\ref{fig:2} (f,g)]. 
Inspection of the harmonics content of the current phase relation confirms that the sign change of the rectification amplitude $\eta$ occurs when $J_1$ reverses its sign [Figs.~\ref{fig:2} (h,i)]. Furthermore, we find that the sign change of $\eta$ does not directly follow the $I_1$ sign change and the maximum of the rectification ($\eta \sim 25 \%$) arises for vortex core positions whose $I_1$, $I_2$, and $J_1$ have comparable size.  
Let us then consider a vortex with odd high-winding number $V^\mathrm{L}_0=3$ assuming a square ($\alpha=1$) shape for the superconducting lead [{Figs.~\ref{fig:2} (j)-(m)}]. For $V^\mathrm{L}_0=3$, we observe that the maximal values of the rectification amplitude occur for the vortex core position that is away from the mirror lines of the superconducting lead [Figs.~\ref{fig:2} (j,k)]. The rectification can reach $30\%$ amplitude and there are multiple positions of the vortex core for which $\eta$ is vanishing [Figs.~\ref{fig:2} (j,k)]. 
Now, when considering the specific case of the current phase relation in the presence of a vortex, since the second harmonic components are usually smaller than the first harmonic, then the condition to become comparable in amplitude is usually met when the first harmonics are vanishingly small or change sign. This happens, e.g.\ nearby $x_0^\mathrm{L} = 22a$ or $7 a$, and in general, nearby the points where the first harmonics become vanishing.
In this context, the appearance of a dip and the peak around specific vortex core positions (see Fig.~\ref{fig:2}  b) results from the simultaneous fact that the first odd-parity harmonic reverses sign and that the amplitudes of the first and second harmonics are of similar magnitude.
\\
The analysis indicates that a key element to maximize the rectification is represented by the search for vortex configurations for which the first harmonics (even and odd-parity) are concomitantly almost vanishing. 
This is a general rule to get large rectification~\cite{ytanakaPRB2022} (see Supplementary Info - Section D). To this aim, one can make an analytical analysis of the first harmonics by considering the direct process of Cooper pairs transfer across the junction as given by~\cite{Josephson62,GolubovRMP2004}
%%%%%%%%%%%%
\begin{align}
    \tilde{I}(\varphi,\bm{r}^\mathrm{L}_{0})
    &\propto \frac{1}{N^\mathrm{L}_{x}N_{y}}\sum_{j_x,j_y} \mathrm{Im} [\Delta_\mathrm{L}(\bm{j},\bm{r}^\mathrm{L}_{0})\Delta^{*}_\mathrm{R}].
\end{align}%
Taking into account the form of the superconducting order parameter we have that:
\begin{align}
    \tilde{I}(\varphi,\bm{r}^\mathrm{L}_{0})
    &\propto \frac{1}{N^\mathrm{L}_{x}N_{y}}\sum_{j_x,j_y}\tilde
    \Theta^\mathrm{L}(\bm{j},\bm{r}^\mathrm{L}_{0})\\
    &\times[\sin\varphi
    \cos\varphi^\mathrm{L}_\mathrm{v}(\bm{j},\bm{r}^\mathrm{L}_{0})+\cos\varphi
    \sin\varphi^\mathrm{L}_\mathrm{v}(\bm{j},\bm{r}^\mathrm{L}_{0})].\notag
\end{align}%
This relation is general and can be applied to any type of vortex phase texture.
Here, from this expression, we can deduce how the nonreciprocal supercurrent directly links to the winding of the phase vortex.
Focusing on the coefficients of $\sin\varphi$ and $\cos\varphi$, the structure of the vortex indeed directly impacts the even and odd-parity components of the first harmonics of the supercurrent.
For convenience and in order to primarily extract the angular dependence of the harmonic content, we neglect the amplitude variation in the vortex core. Hence, the effective first harmonics components $\bar{I}_{1}(\bm{r}^\mathrm{L}_{0})$ and $\bar{J}_{1}(\bm{r}^\mathrm{L}_{0})$ of $\tilde{I}$ are substantially given by 
\begin{align}
    \bar{I}_{1}(\bm{r}^\mathrm{L}_{0})&=\frac{1}{N^\mathrm{L}_{x}N_{y}}\sum_{j_x,j_y}\cos\varphi_\mathrm{L}(\bm{j},\bm{r}^\mathrm{L}_{0}),\\
    \bar{J}_{1}(\bm{r}^\mathrm{L}_{0})&=\frac{1}{N^\mathrm{L}_{x}N_{y}}\sum_{j_x,j_y}\sin\varphi_\mathrm{L}(\bm{j},\bm{r}^\mathrm{L}_{0}).
\end{align}%
We checked that the spatial variation of the superconducting order parameter in the core of the vortex does not alter the qualitative conclusions of the analysis.
\\
We find that their nodal lines can cross in different sites depending on the winding number and aspect ratio of the superconducting lead [Fig.~\ref{fig:3}]. 
In particular, for the case of even winding, $V^\mathrm{L}_0=2$, the breaking of $C_4$ rotational symmetry for a rectangular-shaped superconductor induces a shift of the nodal line of the odd-harmonic $\bar{I}_1$ towards the edge of the superconductor [Fig.~\ref{fig:3} (d),(e)]. 
Then, the crossing with the nodal lines for $\bar{J}_1$ shifts from $(x^\mathrm{L}_{0},y^\mathrm{L}_{0})=(-22a,0)$ to $(x^\mathrm{L}_{0},y^\mathrm{L}_{0})\sim (-22a,\pm 10.5a)$. 
When considering a vortex with winding $V^\mathrm{L}_{0}=3$, due to the higher angular components, the crossings of the vanishing lines for $\bar{I}_1$ are in multiple points within the superconducting domain at $(x^\mathrm{L}_{0},y^\mathrm{L}_{0})\sim (-7a,\pm 7.5a)$ and $(-22a,\pm 7.5a)$ [Fig.~\ref{fig:3} (g,h)]. 
To verify how the spatial profile of the rectification amplitude varies with different vortex states characterized by their winding numbers, we performed numerical computations of the supercurrent in the superconducting junction. These calculations reveal that the rectification pattern exhibits distinct spatial features depending on the vortex winding number.
As illustrated in Fig.~\ref{fig:3} (c), (f), and (i), the pattern of rectification amplitude changes qualitatively with different vortex states. Specifically, for the case where the vortex has a winding number $ V^\mathrm{L}_{0} = 1$, there is a single region where the rectification amplitude vanishes. This region corresponds to the mirror symmetry line with respect to the transformation $y \to -y $. The sign change of the rectification amplitude primarily occurs when crossing this horizontal mirror line at $y = 0 $.
In contrast, for a vortex with winding number $ V^\mathrm{L}_{0} = 2 $, the spatial pattern features two lines where the rectification amplitude changes sign. These sign-change lines are associated with the vortex core's position relative to the interface, which can move from regions far from the interface to those closer to it. As the vortex winding number increases further to $ V^\mathrm{L}_{0} = 3 $, the number of nodal lines—i.e., lines where the rectification amplitude vanishes—expands to four, indicating a more complex spatial structure of the supercurrent distribution influenced by the vortex's winding number.
These findings confirm the qualitative expectation based on the analysis of the first harmonics of the current phase relation.

\noindent
\textbf{Two-vortex configuration.}
\begin{figure}[t!]
    \centering
    \includegraphics[width=8.5cm]{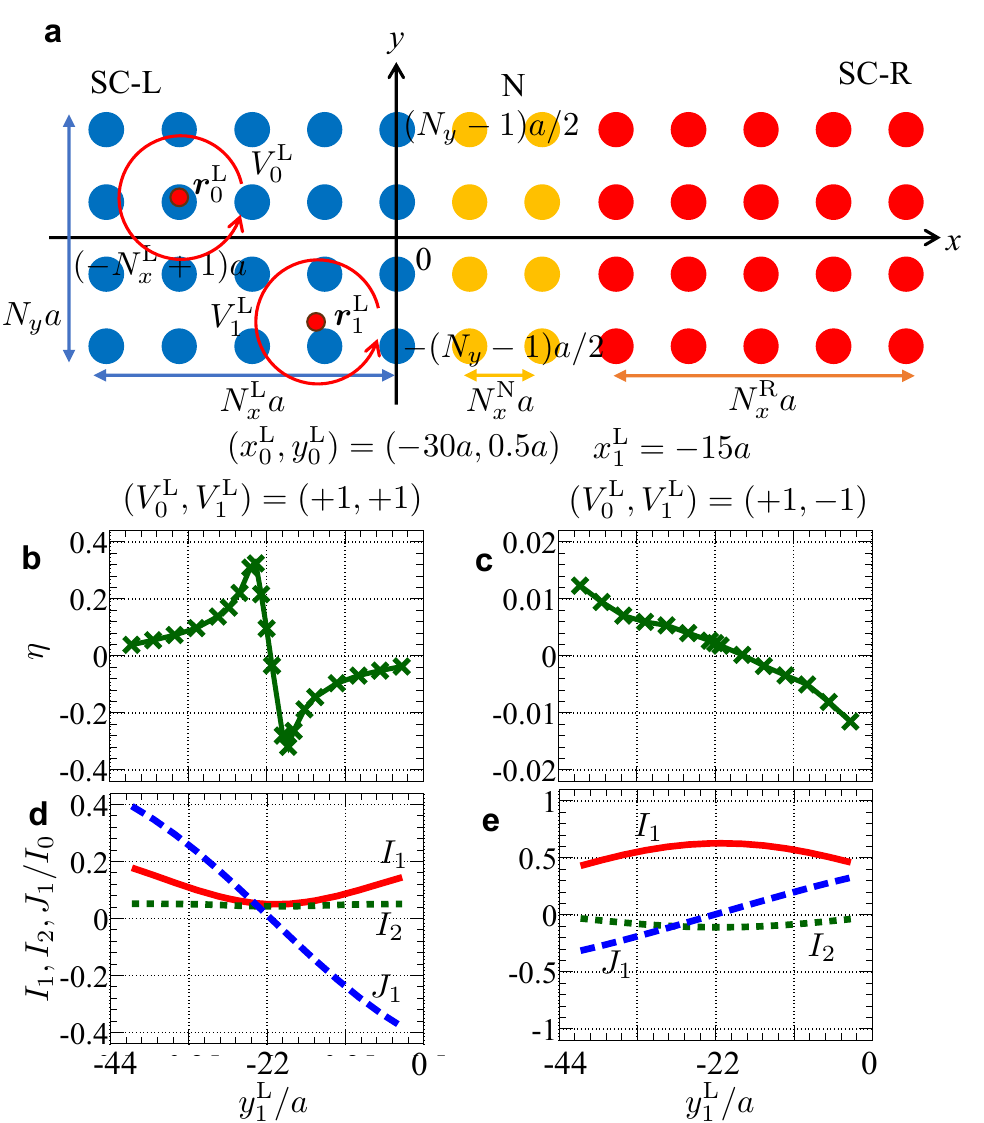}
    \caption{\textbf{Space dependence of rectifications with two vortices.}
    (a) Schematic illustration of two vortices in the left-side superconductor.
    SC-L, N, and SC-R mean the Left-side superconductor (SC), the Normal metal, and the Right-side SC.
    $\bm{r}^\mathrm{L}_{m}=(x^\mathrm{L}_{m},y^\mathrm{L}_{l})$ with $m=1,2$ indicate core position for each vortex.
    $V^\mathrm{L}_{0,1}$ indicate the winding number of each phase vortex.
    $N^\mathrm{L}_{x}$, $N^\mathrm{R}_{x}$, $N^\mathrm{N}_{x}$, $N_{y}$, and $a$ denote the number of sites in the left and right-side SCs, and in the normal metal along the $x$-direction, and number of sites along the $y$-direction, and lattice constant.
    (b,c) Rectification $\eta$ and (d,e) $I_{1}$, $I_{2}$, $J_{1}$ as a function of the literal direction $y^\mathrm{L}_{1}$ for each (b,d) $(V^\mathrm{L}_{0},V^\mathrm{L}_{1})=(+1,+1)$ and (c,e) $(V^\mathrm{L}_{0},V^\mathrm{L}_{1})=(+1,-1)$.
    We set the core positions as $(x^\mathrm{L}_{0},y^\mathrm{L}_{0})=(-30,0.5)$ and $x^\mathrm{L}_{1}=-15$.
    $I_{0}=0.122|\Delta_{0}|(2e/\hbar)$ stands for the maximum Josephson current without any phase vortices in superconductors.
    We select the parameters: $|\Delta_0|=0.02t$ (superconducting energy gap amplitude), $t_\mathrm{int}=0.90$ (transparency at the interface), $N^\mathrm{L}_{x}=N^\mathrm{R}_{x}=45$, $N^\mathrm{N}_{x}=10$, $N_{y}=30$, and $z^\mathrm{L}_{0}=z^\mathrm{L}_{1}=10a$ (each vortex size).}
    \label{fig:4}
\end{figure}%
Next, we show the rectification caused by two vortices in one lead of the superconducting junction.
The pair potential with two vortices in the left-side superconductor is expressed by:
\begin{align}
    \Delta_\mathrm{L}(\bm{j},\bm{r}^\mathrm{L}_{0},\bm{r}^\mathrm{L}_{1})=|\Delta_{0}|\prod^{1}_{m=0}\tilde{\Theta}_{m}(\bm{j},\bm{r}_{m})e^{i\varphi^\mathrm{L}_{\mathrm{v}m}(\bm{j},\bm{r}_{m})},
\end{align}%
with 
\begin{align}
    \tilde{\Theta}_{m}(\bm{j},\bm{r}_{m})=\tanh\left[\frac{|\bm{j}-\bm{r}^\mathrm{L}_{m}|}{z^\mathrm{L}_{m}}\right],
\end{align}%
the size of vortices $z^\mathrm{L}_{m}$, and the core positions $\bm{r}^\mathrm{L}_{m}=(x^\mathrm{L}_{m},y^\mathrm{L}_{m})$ for $m=0,1$.
Each phase vortex is given by
\begin{align}
    \varphi^\mathrm{L}_{\mathrm{v}m}(\bm{j})=V^\mathrm{L}_{m}\arg[(j_{x}a-x^\mathrm{L}_{m})+i(j_{y}a-y^\mathrm{L}_{m})],
\end{align}%
where $V^\mathrm{l}$ is the number of windings for each phase vortex.
\\
We plot the rectification and Josephson components ($I_{1}$, $I_{2}$, and $J_{1}$) as a function of the literal direction $\tilde{y}^\mathrm{L}_{1}$ in a representative case [Fig~\ref{fig:4}].
Then we set the position of one phase vortex at $(x^\mathrm{L}_{0},y^\mathrm{L}_{0})=(-30a,0.5a)$.
We choose each winding number as Fig.~\ref{fig:4} (b,d) $(V^\mathrm{L}_{0},V^\mathrm{L}_{1})=(+1,+1)$ and (c,e) $(V^\mathrm{L}_{0},V^\mathrm{L}_{1})=(+1,-1)$.
For $(V^\mathrm{L}_{0},V^\mathrm{L}_{1})=(+1,+1)$, the rectification is enhanced up to 40\% near $y^\mathrm{L}_{1}=0$ [Fig.~\ref{fig:4} (b)] owing to the comparable Josephson components $|I_{1}|\sim|I_{2}|\sim|J_{1}|$ [Fig.~\ref{fig:4} (d)].
On the other hand, for $(V^\mathrm{L}_{0},V^\mathrm{L}_{1})=(+1,-1)$, because $|I_{1}|$ is larger than $|I_{2}|$ and $|J_{1}|$ [Fig.~\ref{fig:4} (e)], the amplitude of the rectification is small compared with that for $(V^\mathrm{L}_{0},V^\mathrm{L}_{1})=(+1,+1)$ [Fig.~\ref{fig:4} (c)].
Thus, the same sign for winding is favorable for enhancing the rectification.
\begin{figure}[t!]
    \centering
    \includegraphics[width=8.5cm]{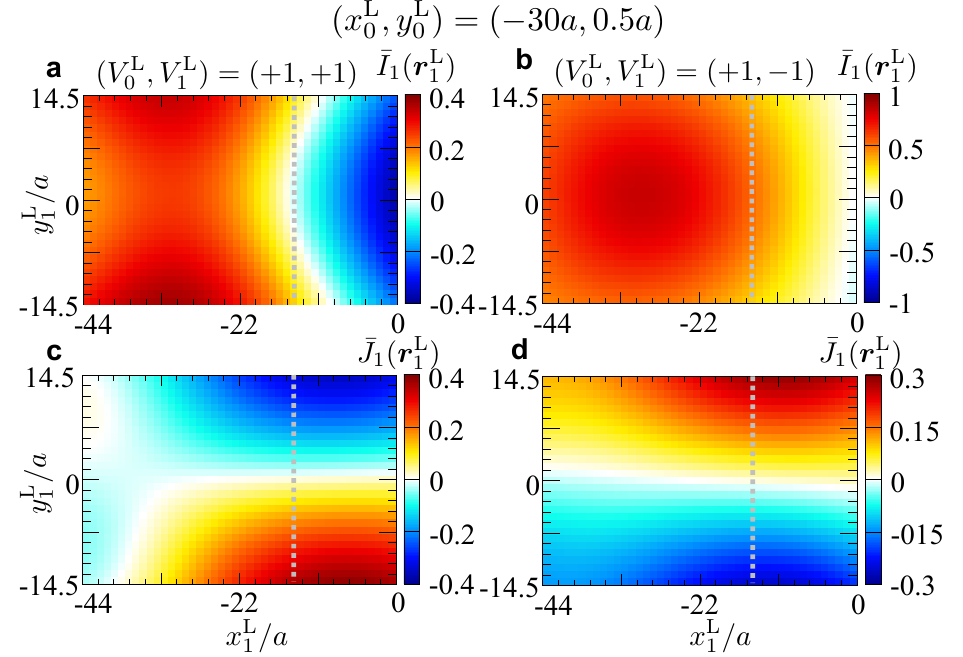}
    \caption{\textbf{Evaluation of the first and second harmonics Josephson current with two vortices.}
    (a,b) $\bar{I}_{1}$ and (c,d) $\bar{J}_{1}$ for the $\bm{r}^\mathrm{L}_{1}=(x^\mathrm{L}_{1},y^\mathrm{L}_{1})$ space at (a,c) $(V^\mathrm{L}_{0},V^\mathrm{L}_{1})=(+1,+1)$ and (b,d) $(V^\mathrm{L}_{0},V^\mathrm{L}_{1})=(+1,-1)$.
    The color bars indicate the amplitude of (a,c) $\bar{I}_1(\bm{r}^\mathrm{L}_{0})$ and (b,d) $\bar{J}_1(\bm{r}^\mathrm{L}_{0})$.
    We set the core positions as $(x^\mathrm{L}_{0},y^\mathrm{L}_{0})=(-30a,0.5a)$ and the vortex size as $z^\mathrm{L}_{1}=10a$ ($N^\mathrm{L}_{x}=N^\mathrm{R}_{x}=45$ and $N_{y}=30$).
    The Gray line indicates the scanning site in Fig.~\ref{fig:4}.}
    \label{fig:5}
\end{figure}%
\\
Based on the discussion in the case of one phase vortex, we can also define $\bar{I}_{1}$ and $\bar{J}_{1}$ in this case.
$\bar{I}_{1}(\bm{r}^\mathrm{L}_{0},\bm{r}^\mathrm{L}_{1})$ and $\bar{J}_{1}(\bm{r}^\mathrm{L}_{0},\bm{r}^\mathrm{L}_{1})$ are given by
\begin{align}
    \bar{I}_{1}(\bm{r}^\mathrm{L}_{0},\bm{r}^\mathrm{L}_{1})&=\frac{1}{N^\mathrm{L}_{x}N_{y}}\sum_{j_x,j_y}\cos\varphi^\mathrm{L}_\mathrm{v}(\bm{j},\bm{r}^\mathrm{L}_{0},\bm{r}^\mathrm{L}_{1}),\\
    \bar{J}_{1}(\bm{r}^\mathrm{L}_{0},\bm{r}^\mathrm{L}_{1})&=\frac{1}{N^\mathrm{L}_{x}N_{y}}\sum_{j_x,j_y}\sin\varphi^\mathrm{L}_\mathrm{v}(\bm{j},\bm{r}^\mathrm{L}_{0},\bm{r}^\mathrm{L}_{1}),
\end{align}%
with $\varphi^\mathrm{L}_{\mathrm{v}}(\bm{j},\bm{r}^\mathrm{L}_{0},\bm{r}^\mathrm{L}_{1})=\varphi^\mathrm{L}_{\mathrm{v}0}(\bm{j},\bm{r}^\mathrm{L}_{0})+\varphi^\mathrm{L}_{\mathrm{v}1}(\bm{j},\bm{r}^\mathrm{L}_{1})$.
Fixing the position of one phase vortex at $(x^\mathrm{L}_{0},y^\mathrm{L}_{0})=(-30a,0.5a)$, we plot $\bar{I}_{1}(\bm{r}^\mathrm{L}_{0},\bm{r}^\mathrm{L}_{1})$ and $\bar{J}_{1}(\bm{r}^\mathrm{L}_{0},\bm{r}^\mathrm{L}_{1})$ for the $\bm{r}^\mathrm{L}_{1}$ space as shown in Fig.~\ref{fig:5}.
For $(V^\mathrm{L}_{0},V^\mathrm{L}_{1})=(+1,+1)$ shown in Fig.~\ref{fig:5} (a,c), since nodal lines appear around $x^\mathrm{L}_{1}\sim -11a$ for $\bar{I}_{1}$ and $y^\mathrm{L}_{1}\sim 0$ for $J_{1}$, the amplitude of $I_{1}$ and $J_{1}$ are small near these lines, respectively.
For $(V^\mathrm{L}_{0},V^\mathrm{L}_{1})=(+1,-1)$, we obtain $\bar{I}_{1}\le 0$ and the nodal line around $y^\mathrm{L}_{1}\sim 0$ for $J_{1}$ [Fig.~\ref{fig:5} (b,d)].
Based on these $\bar{I}_{1}$ and $J_{1}$, $|I_{1}|$ is larger than $I_{1}$ and $J_{1}$ [Fig.~\ref{fig:4} (d)].
Thus, using $\bar{I}_{1}$ and $\bar{J}_{1}$ is generic way to express $I_{1}$ and $J_{1}$.

\noindent\large{\textbf{Conclusions}}\normalsize\\ 
We have demonstrated that nonreciprocal supercurrents are achieved in the presence of high-winding vortex and multiple vortices.
The resulting behavior contains distinct features that can be exploited to distinguish physical configurations with vortices having winding number equal to one, coexistence of vortices and antivortices, or the occurrence of a giant vortex.
In particular, we uncover the spatial profile of the supercurrent rectification with respect to the vortex core position. The rectification pattern depends on the number of windings, with an increase in the winding number producing more complex nodal structures. We find that for $V^\mathrm{L}_\mathrm{0}=1$, a single sign change line exists, while for 2 and 3, multiple lines and nodal points appear, demonstrating how winding number shapes the spatial rectification behavior.
Our findings indicate that the achieved vortex diodes do not exhibit a high rectification efficiency. However, in this context, unlike semiconducting diodes, there are no fundamental reasons to exclude the use of superconducting diodes in superconducting electronics and quantum circuitry, even if their rectification efficiency does not reach $100\%$. 
Recent reports indeed highlighted the potential use of superconducting diodes in diverse applications \cite{Upadhyay2024,Castellani2025,Ingla-Ayno2025}, demonstrating that even with low rectification efficiencies, superconducting diodes can effectively perform functions such as alternating current (AC) to direct current (DC) conversion and rectification.
In particular, one of the primary applications of superconducting diodes involves converting AC to DC at low temperatures to generate stable and adjustable DC bias currents from radiofrequency signals. They achieve this with a rectification efficiency below $50\%$ exploiting vortex dynamics, with the nonreciprocal critical current that arises from the asymmetric expulsion of vortices from the superconducting nanostructure.
In our vortex diode design, instead, we assume that the nonreciprocal supercurrent can be controlled by the vortex position within the junction, without the need to move the vortex itself, where one can achieve rectification amplitudes of the order of $30\%$. 
In this framework, it is worth pointing out that, with respect to the design of vortex diodes and control knobs, vortices can be manipulated (e.g. displaced, introduced, or removed) by magnetic field \cite{Golod2010,Reichhardt_2017,Ma2020}, current \cite{Golod2015,Sok1994,Milosevic2009}, light \cite{Veshchunov2016,Mironov2017}, and mechanical strain \cite{Kremen2016}. 
\begin{figure}[t!]
    \centering
    \includegraphics[width=8.5cm]{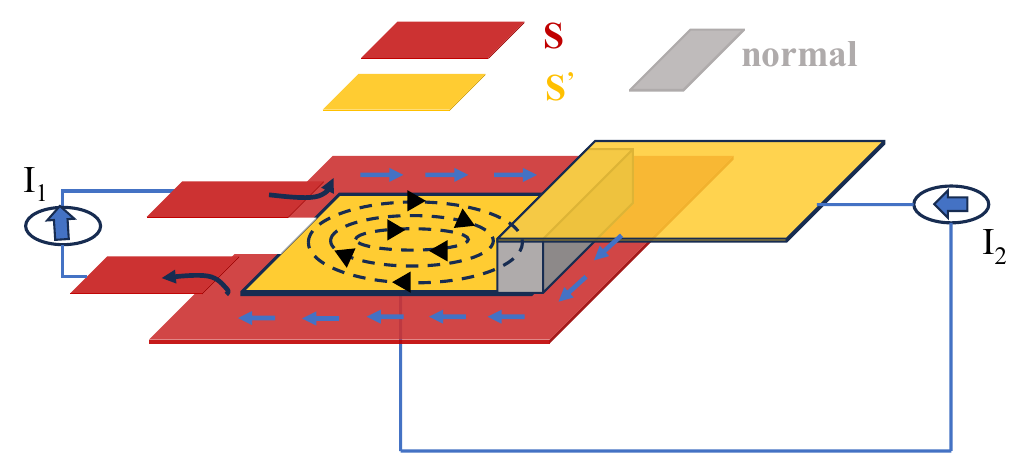}
    \caption{\textbf{A schematic illustration of a physical setup designed to probe nonreciprocal phenomena with high-winding vortices.} 
    An external current source ($I_1$) injects current into a superconductor labeled $S$ (depicted in red). 
    The circulating current influences a nearby proximitized superconductor, $S^{'}$ (shown in orange), leading to the nucleation of an electrically controlled vortex within it \cite{Amundsen2018}. 
    A supercurrent, $I_2$, flows through a weak link based on $S^{'}$ superconducting leads, where an electrically controlled vortex exists on one side of the junction. 
    The nonreciprocal behavior of $I_2$ can be exploited to identify the nature of the induced vortex.}
    \label{fig:6}
\end{figure}%
\\
We would like to discuss how the achieved results depend on system parameters like temperature, junction interface, and disorder. 
Within our model description, the effects of temperature are to substantially reduce the amplitude of the superconducting order parameter. Hence, we do not expect qualitative changes but rather a modification of the amplitude of the supercurrent. 
Regarding the disorder, as shown in Section D of the Supplementary Information, we find that the rectification amplitude is not much altered by the local disorder potential. 
What is more relevant is the junction transparency as it enters to modify the second harmonic component of the supercurrent, as predicted by the Kulik-Omelyanchuk theory, for a superconducting weak link. We have examined the rectification amplitude for a representative vortex configuration in terms of the junction transparency for a representative vortex configuration (see Supplementary Information - Section D). 
Our results confirm that the rectification amplitude gets suppressed when the superconducting junction is brought into the tunneling regime. 
\\
It is also interesting to comment on the possibility of having a coexistence of vortex states and finite momentum pairing, a physical scenario that might lead to nontrivial effects for the superconducting vortex diode. However, it is unlikely that the vortex phase enables the formation of finite momentum pairing. This can happen for large values of the applied magnetic field when a vortex lattice can coexist with a finite momentum pairing of the FFLO type \cite{Shimahara2009}).
This is, however, in a regime of an applied magnetic field, which is beyond the examined cases in our paper because the coexistence occurs close to the upper critical field. Our study refers to a small applied magnetic field where only a few vortices nucleate into the superconductor.
\\
We also would like to mention that our results refer to short superconducting junctions. In short Josephson junctions, the current phase relation's structure and strong coupling can enhance diode efficiency by creating pronounced asymmetries. Instead, long Josephson junctions (LJJs), with their complex phase dynamics and fluxon motion, can manifest different mechanisms for the diode effect. The fluxonium diode, for instance, employs a control line to induce magnetic field asymmetry, though its performance remains not fully tested \cite{Raissi1994,Raissi2003}. Another concept involves a single annular junction with a control line requiring precise flux insertion \cite{Carapella2001}. Recent advances include asymmetric inline LJJs demonstrating sizable superconducting diode effects \cite{Guarcello2024}.
%Remarkably, the configuration for maximal rectification amplitude is not trivially related to the vortex winding at a given core position. 
A suitable platform to observe these effects can be based on Josephson junctions made of nanoislands in the presence of an applied magnetic field~\cite{Kanda2004}. It is known that giant vortices can be induced by a magnetic field for superconductors with coherence length much smaller than the magnetic penetration depth (e.g.\ Pb or Nb based nanostructures)~\cite{Kanda2004}.
In particular, as demonstrated by the theoretical and experimental results in Ref. ~\cite{Kanda2004}, for systems with a size approximately five times the coherence length, the application of a magnetic field on the order of tens of millitesla can induce transitions involving changes in the vortex winding configurations. Our calculations are based on a system size and coherence length that align with this analysis.
Then, starting from a low magnetic field configuration with a vortex having, for instance, winding $V^\mathrm{L}_{0}=1$ nearby the center of the superconducting lead, the transition to a higher winding vortex state by the increase of the magnetic field will be accompanied by a sizable variation of the rectification. Detection of such transitions can be used to probe the high-winding vortex phase.
\\
Along this line, one can also envision a field-free platform~\cite{Amundsen2018} with circular current flow inducing high-winding vortices that concomitantly yield supercurrent rectification (see Fig.~\ref{fig:6}). The described setup provides a potential approach to investigating nonreciprocal phenomena through the manipulation of high-winding vortices. 
By utilizing an external current to induce a vortex in a proximitized superconductor and examining the resulting supercurrent behavior across a weak link, this configuration offers a means to characterize and understand the underlying physics of nonreciprocity in superconducting systems with nontrivial vortex states. 
Such insights could pave the way for advanced superconducting devices with directional control and enhanced functionalities.
Moreover, since a generic phase vortex texture can be expanded in harmonics by employing vortex configurations with different windings, then, in principle, our results provide a toolkit to design the supercurrent rectification for a wide variety of superconducting phase patterns.

\noindent\large{\textbf{Data availability}}\normalsize\\
\noindent {The data that support the findings of this study are available
from the corresponding author upon reasonable request.}

\bibliography{main}

\noindent\large{\textbf{Acknowledgments}}\normalsize\\ 
This work was funded by the EU’s Horizon 2020 Research and Innovation Framework Program under Grant Agreement No.\ 964398 (SUPERGATE), No.\ 101057977 (SPECTRUM), by the PNRR MUR project PE0000023-NQSTI, by the PRIN project 2022A8CJP3 (GAMESQUAD), and by the MAECI project ``ULTRAQMAT".
Y.\,F.\, acknowledges the numerical support from Okayama University.
We thank S.\ Ikegaya and Y.\ Tanaka for valuable discussions.
\\
\noindent\large{\textbf{Author contributions}}\normalsize\\ 
M.C.\ conceived and supervised the project. 
Y.F.\ performed the computations.
The manuscript was
written by Y.F., M.C.\, and C.O.\ with inputs from all the authors.
All authors discussed the results and their implications equally at
all stages. 
\\
\noindent\large{\textbf{Competing interests}}\normalsize\\ 
The authors declare no competing interests.\\
\noindent\textbf{Correspondence} and requests for materials should be addressed to Yuri Fukaya and Mario Cuoco.

\end{document}

%\title{Nonreciprocal supercurrent by vortex phase texture}

\title{Supplementary Information: Supercurrent diode with high winding vortex}

\author{Yuri Fukaya}
\email{fukaya.yuri@spin.cnr.it}
\affiliation{CNR-SPIN, I-84084 Fisciano (SA), Italy, c/o Universit\'a di Salerno, I-84084 Fisciano (SA), Italy}
\affiliation{Faculty of Environmental Life, Natural Science and Technology, Okayama University, 700-8530 Okayama, Japan}

\author{Maria Teresa Mercaldo}
\affiliation{Dipartimento di Fisica ``E. R. Caianiello", Universit\`a di Salerno, IT-84084 Fisciano (SA), Italy}

\author{Daniel Margineda}
%\email{daniel.margineda@nano.cnr.it}
\affiliation{NEST Istituto Nanoscienze-CNR and Scuola Normale Superiore, I-56127, Pisa, Italy}

\author{Alessandro Crippa}
\affiliation{NEST Istituto Nanoscienze-CNR and Scuola Normale Superiore, I-56127, Pisa, Italy}

\author{Elia Strambini}
\affiliation{NEST Istituto Nanoscienze-CNR and Scuola Normale Superiore, I-56127, Pisa, Italy}

\author{Francesco Giazotto}
%\email{francesco.giazotto@sns.it}
\affiliation{NEST Istituto Nanoscienze-CNR and Scuola Normale Superiore, I-56127, Pisa, Italy}

\author{Carmine Ortix}
\affiliation{Dipartimento di Fisica ``E. R. Caianiello", Universit\`a di Salerno, IT-84084 Fisciano (SA), Italy}

\author{Mario Cuoco}
%\email{mario.cuoco@spin.cnr.it}
\affiliation{CNR-SPIN, I-84084 Fisciano (SA), Italy, c/o Universit\'a di Salerno, I-84084 Fisciano (SA), Italy}

\maketitle

% \noindent\large{\textbf{Supplementary Information}}\normalsize\\ 
% \noindent\textbf{A.\ Total matrix of the model Hamiltonian in Josephson junctions}
\section{A.\ Total matrix of the model Hamiltonian in Josephson junctions}
We provide the details of the model Hamiltonian in Josephson junctions.
\begin{eqnarray}
    \hat{H}=\left(
    \begin{array}{cccc|cccc|cccc}
        \hat{u}_\mathrm{L} & \hat{t}_\mathrm{L} & & & & &&& \\
        \hat{t}^{\dagger}_\mathrm{L} & \hat{u}_\mathrm{L} &\ddots & & & && &&&&\\
         & \ddots & \ddots  & \hat{t}_\mathrm{L} & & && &&&\\
         & & \hat{t}^{\dagger}_\mathrm{L} & \hat{u}_\mathrm{L} & \hat{t}_\mathrm{LJ} & && &&&& \\ \hline
         & & & \hat{t}^{\dagger}_\mathrm{LJ} & \hat{u}_\mathrm{N} &\hat{t}_\mathrm{N} && &&&&\\
         & & & & \hat{t}^{\dagger}_\mathrm{N} & \hat{u}_\mathrm{N} & \ddots & &&&&\\
         & & & & & \ddots & \ddots &\hat{t}_\mathrm{N} &&&&\\
         & & & & & & \hat{t}^{\dagger}_\mathrm{N} & \hat{u}_\mathrm{N} & \hat{t}_\mathrm{RJ} &&& \\ \hline
         &&&&&& & \hat{t}^{\dagger}_\mathrm{RJ} & \hat{u}_\mathrm{R} & \hat{t}_\mathrm{R} && \\
         &&&&&& & & \hat{t}^{\dagger}_\mathrm{R} & \hat{u}_\mathrm{R} & \ddots & \\
         &&&&&& & & & \ddots & \ddots & \hat{t}_\mathrm{R} \\
         &&&&&& & & & & \hat{t}^{\dagger}_\mathrm{R} &\hat{u}_\mathrm{R}
    \end{array}\right).
\end{eqnarray}%
in the basis
\begin{eqnarray}
    C^{\dagger}=&&[c^{\dagger}_{-N^\mathrm{L}_x+1,-(N_y-1)/2,\uparrow},c_{-N^\mathrm{L}_x+1,-(N_y-1)/2,\downarrow},\cdots c^{\dagger}_{-N^\mathrm{L}_x+1,(N_y-1)/2,\uparrow},c_{-N^\mathrm{L}_x+1,(N_y-1)/2,\downarrow},\notag\\
    &&\cdots,c^{\dagger}_{N^\mathrm{N}_{x}+N^\mathrm{R}_{x},-(N_y-1)/2,\uparrow},c_{N^\mathrm{N}_{x}+N^\mathrm{R}_{x},-(N_y-1)/2,\downarrow},\cdots, c^{\dagger}_{N^\mathrm{N}_{x}+N^\mathrm{R}_{x},(N_y-1)/2,\uparrow},c_{N^\mathrm{N}_{x}+N^\mathrm{R}_{x},(N_y-1)/2,\downarrow}].
\end{eqnarray}
Each sector in the matrix is given by
\begin{eqnarray}
    \hat{u}_\mathrm{L}&&
    =\left(
    \begin{array}{cc|cc|cc|cc}
        -\varepsilon & \Delta_\mathrm{L} & -t & && && \\
        \Delta^{*}_\mathrm{L} & \varepsilon & & t && &&\\ \hline
        -t & & -\varepsilon & \Delta_\mathrm{L} & \ddots& && \\
          & t & \Delta^{*}_\mathrm{L} & \varepsilon && \ddots &&\\ \hline
         & & \ddots & & \ddots & & -t & \\
         & & & \ddots & & \ddots && t \\ \hline
         && && -t & & -\varepsilon & \Delta_\mathrm{L} \\
         && && & t & \Delta^{*}_\mathrm{L} & \varepsilon
    \end{array}\right),
\end{eqnarray}
\begin{eqnarray}
    \hat{t}_\mathrm{L}&&=
    \left(
    \begin{array}{cc|cc|cc}
        -t & && && \\
         & t && && \\ \hline
        && \ddots & && \\
        && & \ddots && \\ \hline
        && && -t & \\
        && && & t
    \end{array}
    \right),
\end{eqnarray}%
\begin{eqnarray}
    \hat{u}_\mathrm{N}&&=
    \left(
    \begin{array}{cc|cc|cc|cc}
        -\varepsilon &  & -t & && && \\
         & \varepsilon & & t && &&\\ \hline
        -t & & -\varepsilon &  & \ddots& && \\
          & t &  & \varepsilon && \ddots &&\\ \hline
         & & \ddots & & \ddots & & -t & \\
         & & & \ddots & & \ddots && t \\ \hline
         && && -t & & -\varepsilon &  \\
         && && & t &  & \varepsilon
    \end{array}\right),
\end{eqnarray}%
\begin{eqnarray}
    \hat{t}_\mathrm{N}&&=\left(
    \begin{array}{cc|cc|cc}
        -t & && && \\
         & t && && \\ \hline
        && \ddots & && \\
        && & \ddots && \\ \hline
        && && -t & \\
        && && & t
    \end{array}
    \right),
\end{eqnarray}%
\begin{eqnarray}
    \hat{u}_\mathrm{R}&&=\left(
    \begin{array}{cc|cc|cc|cc}
        -\varepsilon & \Delta_\mathrm{R} & -t & && && \\
        \Delta^{*}_\mathrm{R} & \varepsilon & & t && &&\\ \hline
        -t & & -\varepsilon & \Delta_\mathrm{R} & \ddots& && \\
          & t & \Delta^{*}_\mathrm{R} & \varepsilon && \ddots &&\\ \hline
         & & \ddots & & \ddots & & -t & \\
         & & & \ddots & & \ddots && t \\ \hline
         && && -t & & -\varepsilon & \Delta_\mathrm{R} \\
         && && & t & \Delta^{*}_\mathrm{R} & \varepsilon
    \end{array}\right),
\end{eqnarray}%
\begin{eqnarray}
    \hat{t}_\mathrm{R}&&=\left(
    \begin{array}{cc|cc|cc}
        -t & && && \\
         & t && && \\ \hline
        && \ddots & && \\
        && & \ddots && \\ \hline
        && && -t & \\
        && && & t
    \end{array}
    \right),
\end{eqnarray}%
\begin{eqnarray}
    \hat{t}_\mathrm{LJ}&&=\hat{t}_\mathrm{RJ}=t_\mathrm{int}\left(
    \begin{array}{cc|cc|cc}
        -t & && && \\
         & t && && \\ \hline
        && \ddots & && \\
        && & \ddots && \\ \hline
        && && -t & \\
        && && & t
    \end{array}
    \right),
\end{eqnarray}%
where $t$, $\varepsilon=-0.25t$, $\Delta_\mathrm{L,R}$, and $t_\mathrm{int}$ are the nearest-neighbor hopping integral, onsite energy, the pair potential, and the ratio of the transparency in junctions, respectively.
\newline

\noindent
% \textbf{B.\ Phase vortex in real space}
\section{B.\ Phase vortex in real space}
We present the model and the methodology employed to calculate the Josephson current of the superconducting weak-link in the presence of vortices with different winding numbers.

\begin{figure}[t!]
    \centering
    \includegraphics[width=8.5cm]{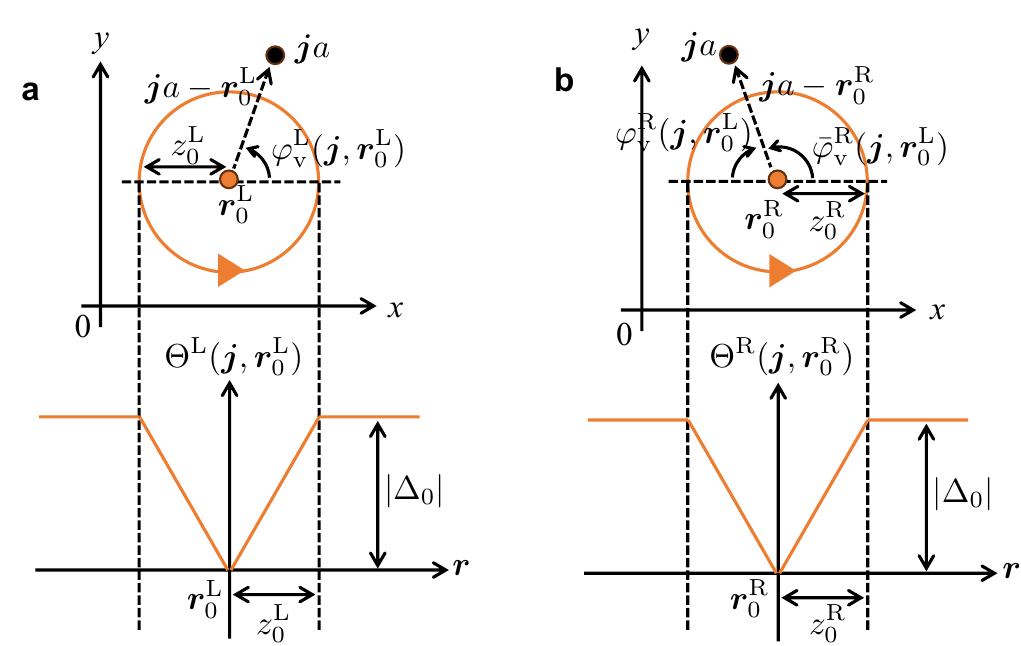}
    \caption{
    (a)(b) Phase of the vortex $\varphi^\mathrm{X}_\mathrm{v}(\bm{j},\bm{r}^\mathrm{L}_{0})$ and corresponding superconducting order parameter $\Theta^\mathrm{X}(\bm{j},\bm{r}^\mathrm{L}_{0})$ in the real space with (a) X=L and (b) X=R.
    We also show the corresponding notations for the phase vortex for each side. $z^\mathrm{X}_0$ indicates the size of the vortex core wheareas the amplitude superconducting order parameter vanishes.}
    \label{figa:1}
\end{figure}

The phase of the superconducting order parameter associated with the vortex configuration is represented in Fig.~\ref{figa:1} (b).
At the site $\bm{j}=(j_{x},j_{y})$, the phase vortex in the left-side superconductor is given by
\begin{align}
    \varphi^\mathrm{L}_\mathrm{v}(\bm{j},\bm{r}^\mathrm{L}_{0})=V^\mathrm{L}_{0}\arg[(j_{x}a-x^\mathrm{L}_{0})+i(j_{y}a-y^\mathrm{L}_{0})],
\end{align}%
with the position of the core $\bm{r}^\mathrm{L}_{0}=(x^\mathrm{L}_{0},y^\mathrm{L}_{0})$.
In a similar way, one can introduce the phase vortex at the core $\bm{r}^\mathrm{R}_{0}=(x^\mathrm{R}_{x},y^\mathrm{R}_{0})$ in the right-side superconductor as
\begin{align}
    \varphi^\mathrm{R}_\mathrm{v}(\bm{j},\bm{r}^\mathrm{R}_{0})&=V^\mathrm{R}_{0}\pi-\bar{\varphi}^\mathrm{R}_\mathrm{v}(\bm{j},\bm{r}^\mathrm{R}_{0}),\\
    \bar{\varphi}^\mathrm{R}_\mathrm{v}(\bm{j},\bm{r}^\mathrm{R}_{0})&=V^\mathrm{R}_{0}\arg[(j_{x}a-x^\mathrm{R}_{0})+i(j_{y}a-y^\mathrm{R}_{0})],
\end{align}%
due to the mirror symmetry in the $yz$-plane.
Here, $V^\mathrm{X}_{0}$ denotes the vortex winding.

The pair potential is described as
\begin{align}
    \Delta_\mathrm{X}=
    \begin{cases}
        \Theta(\bm{j},\bm{r}^\mathrm{X}_{0})
        e^{i\varphi^\mathrm{X}_\mathrm{v}(\bm{j},\bm{r}^\mathrm{X}_{0})}& \text{with phase vortex} \\
        |\Delta_{0}| & \text{without phase vortex}
    \end{cases},
\end{align}%
with the energy gap amplitude $|\Delta_{0}|$.
In the former case, the energy gap amplitude $\Theta(\bm{j},\bm{r}^\mathrm{X}_{0}))$ is modified by the site:
\begin{align}
    \Theta^\mathrm{X}(\bm{j},\bm{r}^\mathrm{X}_{0})&=|\Delta_{0}|\tilde\Theta^\mathrm{X}(\bm{j},\bm{r}^\mathrm{X}_{0}),\\
    \tilde\Theta^\mathrm{X}(\bm{j},\bm{r}^\mathrm{X}_{0})&=\tanh\left[\frac{|\bm{j}a-\bm{r}^\mathrm{X}_{0}|}{z^\mathrm{X}_0}\right],
\end{align}%
with the size of the vortex $z^\mathrm{X}_{0}$.
We set the parameters as $\varepsilon=-0.25t$, $|\Delta_{0}|=0.02t$, and $t_\mathrm{int}=0.90$.
In this study, we choose the size as $N^\mathrm{N}_{x}=10$, $N^\mathrm{R}_{x}=N^\mathrm{L}_{x}=\alpha N_{y}$, $N_{y}=30$ with the aspect ratio $\alpha$, and $z^\mathrm{L}_{0}=10a$.

\begin{figure}[t!]
    \centering
    \includegraphics[width=14.5cm]{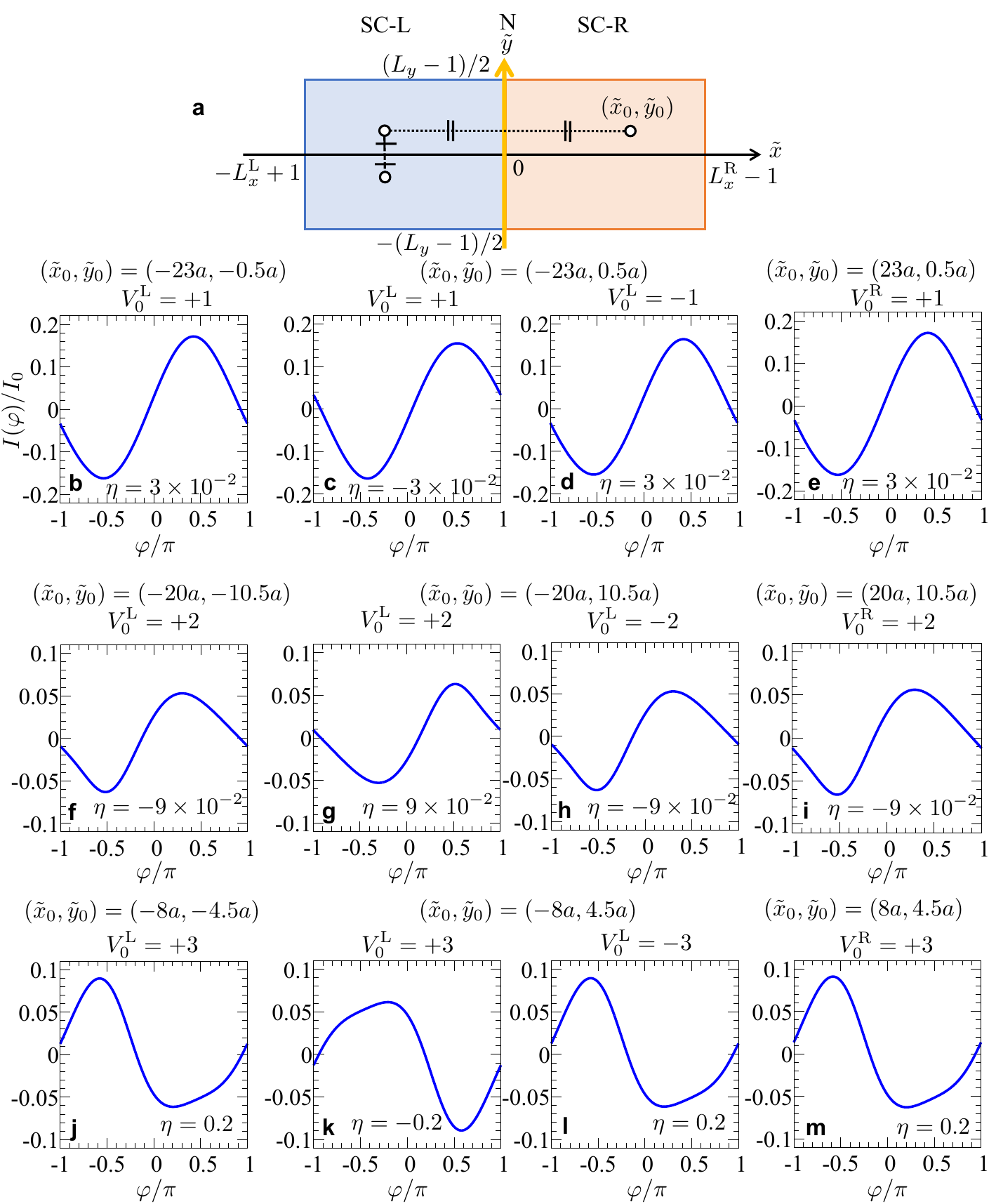}
    \caption{(a) Image of the coordinate to show the symmetry of the current phase relation with a phase vortex.
    We ignore the normal metal (N) region and we focus on the superconductor (SC) regime in this coordinate.
    (b)-(m) Current phase relation for the winding number and the core position for (b)(c) $V^\mathrm{L}_{0}=1$, (d) $V^\mathrm{L}_{0}=-1$, and (e) $V^\mathrm{R}_{0}=1$, (f)(g) $V^\mathrm{L}_{0}=2$, (h) $V^\mathrm{L}_{0}=-2$, and (i) $V^\mathrm{R}_{0}=3$, and (j)(k) $V^\mathrm{L}_{0}=3$, (l) $V^\mathrm{L}_{0}=-3$, and (m) $V^\mathrm{R}_{0}=3$.
    We choose the position of the vortex core at (b) $(\tilde{x}_{0},\tilde{y}_{0})=(-23a,-0.5a)$, (c)(d) $(-23a,0.5a)$, (e) $(23a,0.5a)$, (f) $(-20a,-10.5a)$, (g)(h) $(-20a,10.5a)$, (i) $(20a,10.5a)$, (j) $(-8a,-4.5a)$, (k)(l) $(-8a,4.5a)$, (m) $(8a,4.5a)$.
    System size: (b)-(i) $N^\mathrm{L}_{x}=N^\mathrm{R}_{x}=45$, $N^\mathrm{N}_{x}=10$, and $N_y=30$, and (j)-(m) $N^\mathrm{L}_{x}=N^\mathrm{R}_{x}=30$, $N^\mathrm{N}_{x}=10$, and $N_y=30$. 
    Parameters: $\varepsilon=-0.25t$, $|\Delta_{0}|=0.02t$, $t_\mathrm{int}=0.90$, $z^\mathrm{L}_{0}=z^\mathrm{R}_{0}=10a$, and $I_{0}=0.012|\Delta_{0}|$.}
    \label{CPR_symm}
\end{figure}

We demonstrate the current phase relation $I(\varphi)$ and the corresponding rectification amplitude $\eta$ for each winding number at the core position $(\tilde{x}_{0},\tilde{y}_{0})$ in Fig.~\ref{CPR_symm}.
Then we choose the coordinate as shown in Fig.~\ref{CPR_symm} (a) and we select the mirror-symmetric core position for each winding number in Fig.~\ref{CPR_symm} (b)-(m).
In general, one can show that 
the following relations for the rectification amplitude $\eta$ holds when considering vortex (arbitrary winding amplitude $V_0>0$) or antivortex ($V_0<0$) configurations with a given position of the core $(\tilde{x}_{0},\tilde{y}_{0})$ and mirror related core positions:
\begin{align}
\eta(\tilde{x}_{0},\tilde{y}_{0},V_{0})=-\eta(-\tilde{x}_{0},\tilde{y}_{0},V_{0}),\label{e_x0} \\
\eta(\tilde{x}_{0},\tilde{y}_{0},V_{0})=-\eta(\tilde{x}_{0},-\tilde{y}_{0},V_{0}),\label{e_y0}\\    \eta(\tilde{x}_{0},\tilde{y}_{0},V_{0})=-\eta(\tilde{x}_{0},\tilde{y}_{0},-V_{0}) \,.
\end{align}%
Therefore, regarding a combination of mirror and time-reversal symmetries, one can also obtain that a change from vortex to antivortex and of the vortex core position does not alter the sign and amplitude of the rectification:
\begin{align}
    \eta(\tilde{x}_{0},\tilde{y}_{0},V_{0})
    =\eta(-\tilde{x}_{0},\tilde{y}_{0},-V_{0}).
\end{align}%
\newline

% \noindent\textbf{C.\ Current phase relations in the presence of a single vortex}
\section{C.\ Current phase relations in the presence of a single vortex}
\begin{figure}[t!]
    \centering
    \includegraphics[width=8.5cm]{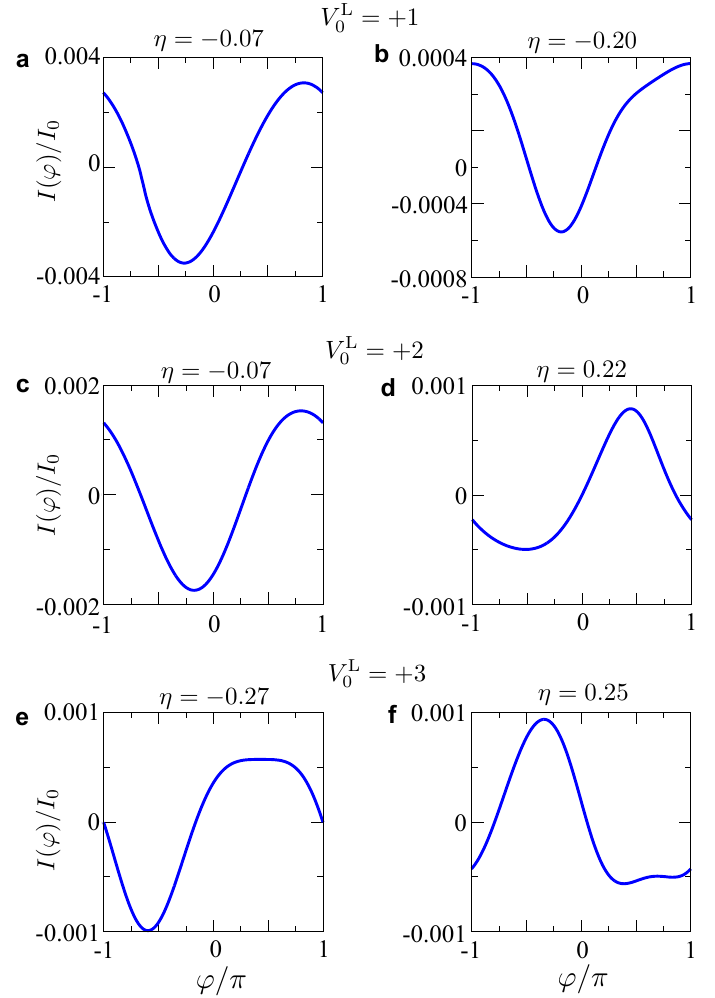}
    \caption{Current phase relation for (a,b) $V^\mathrm{L}_{0}=+1$, (c,d) $+2$, and (e,f) $+3$.
    We choose the position of the vortex core as (a) $(x^\mathrm{L}_{0},y^\mathrm{L}_{0})=(-23a,3.5a)$, (b) $(-20a,0.5a)$, (c) $(-23a,10.5a)$, (d) $(19a,10.5a)$, (e) $(-6a,10.5a)$, and (f) $(-34a,10.5a)$.
    Aspect ratio is (a)-(d) $\alpha=2/3$ and (e,f) $\alpha=1$.
    We also calculate the rectification: (a) $\eta=-0.07$, (b) $-0.20$, (c) $-0.07$, (d) $0.22$, (e) $-0.27$, and (f) $0.25$.
    Parameters: $\varepsilon=-0.25t$, $|\Delta_{0}|=0.02t$, $t_\mathrm{int}=0.90$, $N_{y}=30$, $z^\mathrm{L}_{0}=10a$, and $I_{0}=0.012|\Delta_{0}|$.}
    \label{figa:2}
\end{figure}%
We show the representative cases of the current phase relation corresponding to the physical configurations in the main text.
Figure \ref{figa:2} presents the current phase relation for Fig.~\ref{figa:2} (a)(b) $V^\mathrm{L}_{0}=1$, Fig.~\ref{figa:2} (c)(d) $V^\mathrm{L}_{0}=2$, and Fig.~\ref{figa:2} (e)(f) $V^\mathrm{L}_{0}=3$ at the position of a vortex core (a) $(x^\mathrm{L}_{0},y^\mathrm{L}_{0})=(-23a,3.5a)$, (b) $(-20a,0.5a)$, (c) $(-23a,10.5a)$, (d) $(19a,10.5a)$, (e) $(-6a,10.5a)$, and (f) $(-34a,10.5a)$. 
We obtain the nonreciprocal supercurrents caused by a phase vortex and these rectifications are Fig.~\ref{figa:2} (a) $\eta=-0.07$, Fig.~\ref{figa:2} (b) $-0.20$, Fig.~\ref{figa:2} (c) $-0.07$, Fig.~\ref{figa:2} (d) $0.22$, Fig.~\ref{figa:2} (e) $-0.27$, and Fig.~\ref{figa:2} (f) $0.25$, respectively.
\newline

\begin{figure}[t!]
    \centering
    \includegraphics[width=9.5cm]{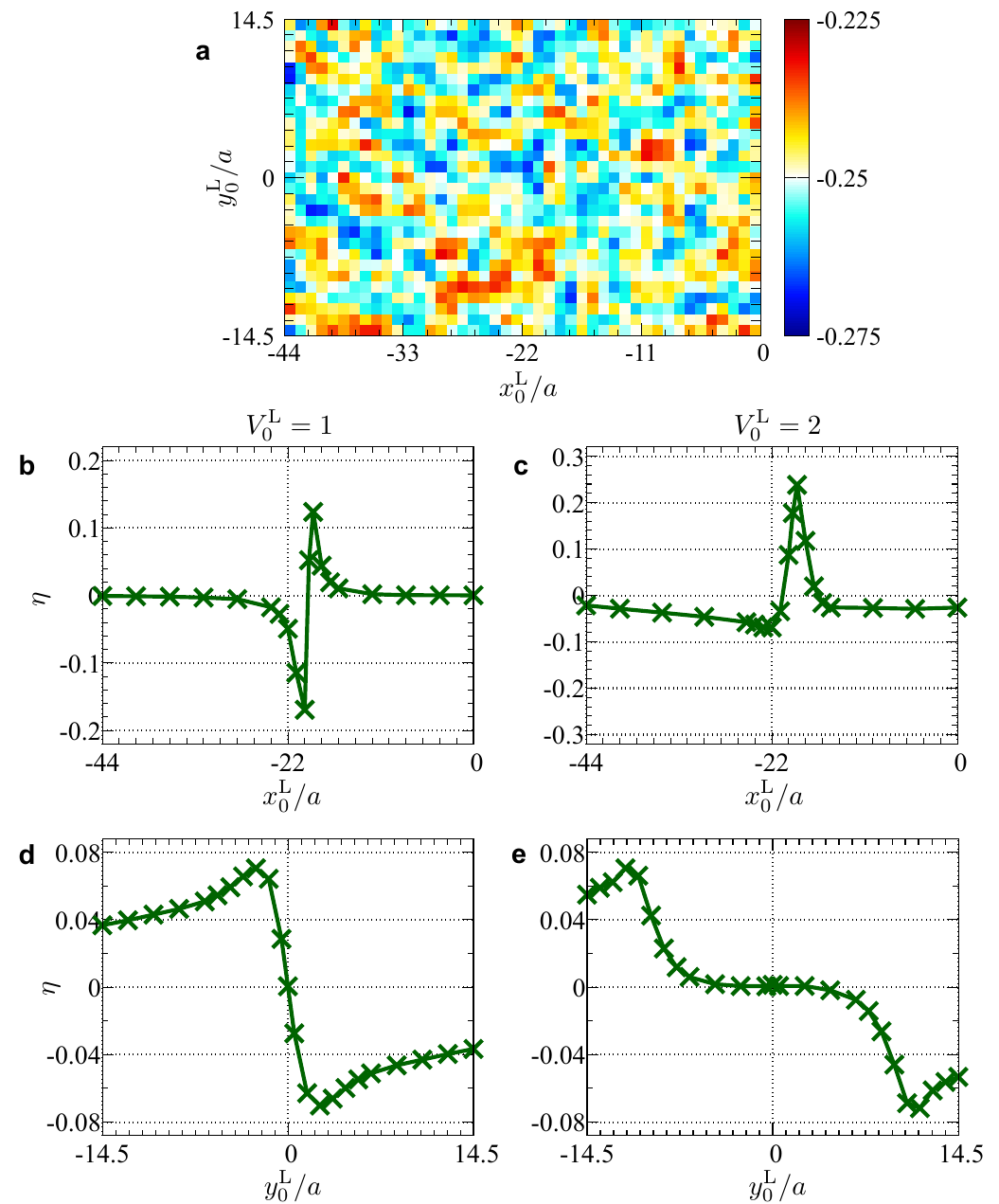}
    \caption{(a) On-site energy $\varepsilon'/t$ distribution in left-side superconductor. 
    We adopt the random number $-0.5\le r\le 0.5$ for each site, and we plot as $\varepsilon'=\varepsilon+0.2|\varepsilon|r$ with the original onsite energy $\varepsilon=-0.25t$.
    (b)-(e) Rectification as a function of (b)(c) $x^\mathrm{L}_{0}$ at (b) $y^\mathrm{L}_{0}=0.5$ and (c) $y^\mathrm{L}_{0}=10.5$, and (d)(e) $y^\mathrm{L}_{0}$ at $x^\mathrm{L}_{0}=-23$.
    We choose the winding number as (b)(d) $V^\mathrm{L}_{0}=1$ and (c)(e) $V^\mathrm{L}_{0}=2$.
    Parameters: $|\Delta_0|=0.02t$, $t_\mathrm{int}=0.90$, $N^\mathrm{L}_{x}=N^\mathrm{R}_{x}=45$, $N^\mathrm{N}_{x}=10$, $N_{y}=30$, and $z^\mathrm{L}_{0}=10a$.}
    \label{figa_add1}
\end{figure}

\begin{figure}
    \centering
    \includegraphics[width=6.5cm]{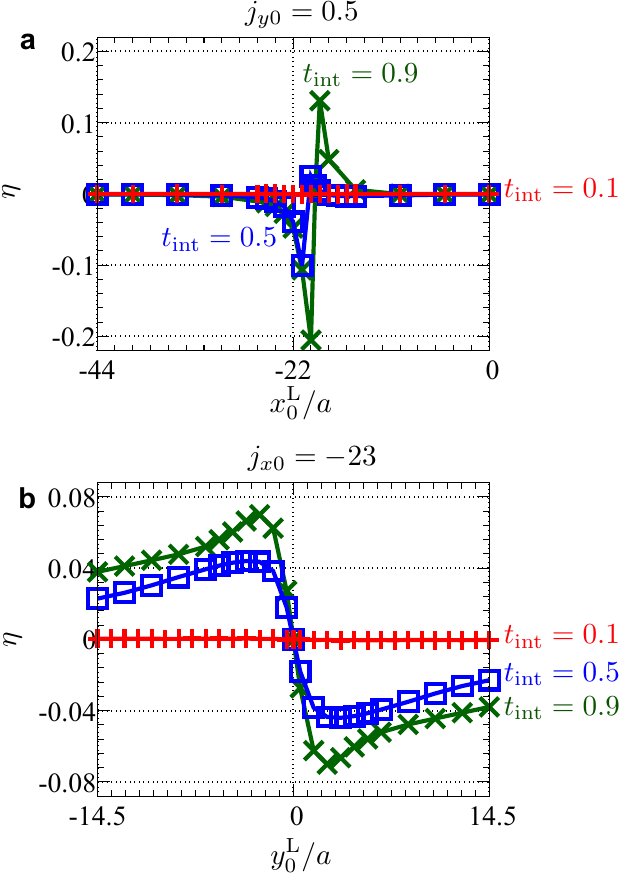}
    \caption{Rectification as a function of (a) $x^\mathrm{L}_{0}$ at $y^\mathrm{L}_{0}=0.5$ and (b) $y^\mathrm{L}_{0}$ at $x^\mathrm{L}_{0}=-23$.
    Green, blue, and red lines indicate at $t_\mathrm{int}=0.9$, $0.5$, and $0.1$, respectively.
    The green line is the same as in Fig.\ 2(b) and Fig.\ 2 (c) of the main text.
    Parameters: $|\Delta_0|=0.02t$, $t_\mathrm{int}=0.90$, $N^\mathrm{L}_{x}=N^\mathrm{R}_{x}=45$, $N^\mathrm{N}_{x}=10$, $N_{y}=30$, $V^\mathrm{L}_0=1$, and $z^\mathrm{L}_{0}=10a$.}
    \label{figa_add2}
\end{figure}

% \noindent\textbf{D.\ Rectification amplitude: role of disorder and junction transparency}
\section{D.\ Rectification amplitude: role of disorder and junction transparency}
For real-device factors, we perform extra simulations to investigate the role of disorder and junction transparency. 
Within our model description, we notice that the effects of temperature are to substantially reduce the amplitude of the superconducting order parameter. 
Hence, we do not expect qualitative changes but rather a modification of the amplitude of the supercurrent. 
In the case of the disorder, as shown in Figure \ref{figa_add1}, we find that the rectification amplitude is not much altered by the local disorder potential. 
There might be fluctuations in the amplitude of the rectification as a function of the disorder strength. These aspects go beyond the scope of the paper.
What is more relevant is the junction transparency as it enters to modify the high-harmonic component of the supercurrent as predicted by the Kulik-Omelyanchuk theory for a superconducting weak link. 
In Figure \ref{figa_add2}, we present the rectification amplitude for a representative vortex configuration. 
It demonstrates that the rectification amplitude is suppressed when the superconducting junction is brought into a tunneling regime.
\newline

% \noindent\textbf{E.\ Condition to maximize the amplitude of the supercurrent rectification}
\section{E.\ Condition to maximize the amplitude of the supercurrent rectification}
\begin{figure}[t!]
    \centering
    \includegraphics[width=8.6cm]{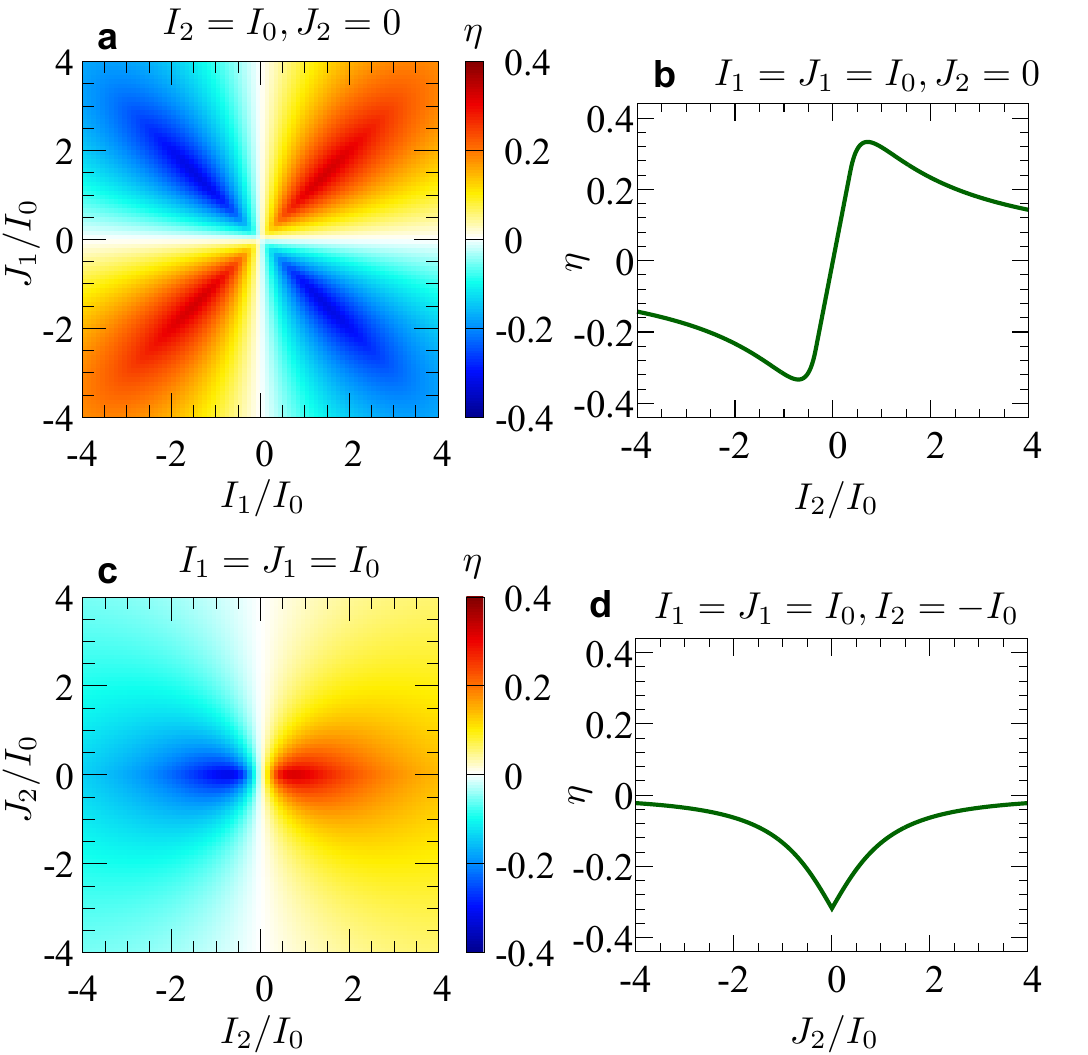}
    \caption{Rectification $\eta$ of $I(\varphi)=I_{1}\sin\varphi+I_{2}\sin 2\varphi+J_{1}\cos\varphi+J_{2}\cos2\varphi$ (a) for $(I_{1},J_{1})$ space and (b) as a function of $I_{2}$.
    We fix the parameters as (a) $(I_{2},J_{2})=(I_{0},0)$, (b) $(I_{1},J_{1},J_{2})=(I_{0},I_{0},0)$, (c) $(I_{1},J_{1})=(I_{0},I_{0})$, and (d) $(I_{1},J_{1},I_{2})=(I_{0},I_{0},-I_{0})$.}
    \label{figa:3}
\end{figure}%
We calculate the optimal conditions to maximize the supercurrent rectification starting from a generic current phase relation that includes the first harmonics (even and odd parity) and the second harmonic (odd parity).
Let us then assume the following current phase relation:
\begin{align*}
    I(\varphi)=I_{1}\sin\varphi+I_{2}\sin 2\varphi+J_{1}\cos\varphi+J_{2}\cos2\varphi.
\end{align*}%
In order to find the optimal regimes for maximal rectification, one can evaluate $\eta$ as a function of the amplitudes $I_{1}$, $I_{2}$, and $J_{1}$.
First, we calculate $\eta$ for $(I_{1},J_{1})$ at $(I_{2},J_{2})=(I_{0},0)$ in Fig.~\ref{figa:3} (a).
The maximum value of $|\eta|$ appears around $|I_{1}|=|J_{1}|\sim I_{0}$.
Based on this point, we also calculate $\eta$ as a function of $I_{2}$ at $(I_{1},J_{1},J_{2})=(I_{0},I_{0},0)$ [Fig.~\ref{figa:3} (b)].
The amplitude of $\eta$ is enhanced around $I_{2}=\pm 0.5I_{0}$.
Likewise, to show the role of $J_{2}$, we calculate $\eta$ for $(I_{2},J_{2})$ at $(I_{1},J_{1}=(I_{0},I_{0})$ in Fig.~\ref{figa:3} (c).
$\eta$ is maximized around $J_{2}\sim 0$ for $|I_{2}|\sim I_{0}$.
We also calculate $\eta$ for $J_{2}$ at $(I_{1},J_{1},I_{2})=(I_{0},I_{0},-I_{0})$ [Fig.~\ref{figa:3} (d)].
The maximum $|\eta|$ appears at $I_{2}=0$.
Thus, both $|I_{1}|$ and $|J_{1}|$ should be comparable with $|I_{2}|$, and $|J_{2}|$ should be minimized to maximize the rectification in Josephson current.
\newline

% \noindent\textbf{F.\ Evaluation of the first-harmonic component of supercurrent}
\section{F.\ Evaluation of the first-harmonic component of supercurrent}
We provide how we obtain the evaluation of the first-harmonic component of the supercurrent.
The first-harmonic components of the Josephson current at a site $\bm{j}$ can be estimated by considering the direct Cooper pair tunneling process as
\begin{align}
    \tilde{J}(\varphi;\bm{j},\bm{r}^\mathrm{L}_{0})
    &\propto \mathrm{Im} [\Delta_\mathrm{L}(\bm{j},\bm{r}^\mathrm{L}_{0})\Delta^{*}_\mathrm{R}],
\end{align}%
where $\bm{r}^\mathrm{L}_{0}$ is the position of a vortex core in the left-side superconductor.
By expanding the expression for $\tilde{J}(\phi)$, we obtain
\begin{align}
\tilde{J}(\varphi;\bm{j},\bm{r}^\mathrm{L}_{0})&\propto \mathrm{Im} \left[ |\Delta_{0}|^{2}\tanh\left[\frac{|\bm{j}a-\bm{r}^\mathrm{L}_{0}|}{z^\mathrm{L}_{0}}\right]e^{i(\varphi_\mathrm{L}-\varphi_\mathrm{R})}e^{i\tilde\varphi^\mathrm{L}_\mathrm{v}(\bm{j},\bm{r}^\mathrm{L}_{0})}\right]\notag\\
    &=|\Delta_{0}|\sin\varphi
    \tanh\left[\frac{|\bm{j}a-\bm{r}^\mathrm{L}_{0}|}{z^\mathrm{L}_{0}}\right]
    \cos\tilde\varphi^\mathrm{L}_\mathrm{v}(\bm{j},\bm{r}^\mathrm{L}_{0})\notag\\
    &+|\Delta_{0}|\cos\varphi\tanh\left[\frac{|\bm{j}a-\bm{r}^\mathrm{L}_{0}|}{z^\mathrm{L}_{0}}\right]\sin\tilde\varphi^\mathrm{L}_\mathrm{v}(\bm{j},\bm{r}^\mathrm{L}_{0}),
\end{align}%
with $\varphi=\varphi_\mathrm{L}-\varphi_\mathrm{R}$.
The resulting first-harmonic Josephson current can be described by the spatial average of $\tilde{J}(\varphi;\bm{j},\bm{r}^\mathrm{L}_{0})$ over the number of sites in the superconducting leads:
\begin{align}
    \Tilde{I}(\varphi;\bm{r}^\mathrm{L}_{0})
    &=\frac{1}{N^\mathrm{L}_{x}N_{y}}\sum_{j_x,j_y}\tilde{J}(\bm{j},\bm{r}^\mathrm{L}_{0})\\
    &\propto\sin\varphi
    \frac{1}{N^\mathrm{L}_{x}N_{y}}\sum_{j_x,j_y}\tanh\left[\frac{|\bm{j}a-\bm{r}^\mathrm{L}_{0}|}{z^\mathrm{L}_{0}}\right]
    \cos\tilde\varphi^\mathrm{L}_\mathrm{v}(\bm{j},\bm{r}^\mathrm{L}_{0})\notag\\
    &+\cos\varphi\frac{1}{N^\mathrm{L}_{x}N_{y}}\sum_{j_x,j_y}\tanh\left[\frac{|\bm{j}a-\bm{r}^\mathrm{L}_{0}|}{z^\mathrm{L}_{0}}\right]\sin\tilde\varphi^\mathrm{L}_\mathrm{v}(\bm{j},\bm{r}^\mathrm{L}_{0}).\notag
\end{align}% 
Based on this analysis, we get that
\begin{align}
    \tilde{I}(\varphi;\bm{r}^\mathrm{L}_{0})&\propto F_\mathrm{c}(\bm{r}^\mathrm{L}_{0})\sin\varphi
    +F_\mathrm{s}(\bm{r}^\mathrm{L}_{0})\cos\varphi, \label{sm_J_r0_discussion}
\end{align}%
with
\begin{align}
    F_\mathrm{c1}(\bm{r}^\mathrm{L}_{0})&=\frac{1}{N^\mathrm{L}_{x}N_{y}}\sum_{j_x,j_y}\tanh\left[\frac{|\bm{j}a-\bm{r}^\mathrm{L}_{0}|}{z^\mathrm{L}_{0}}\right]\cos\tilde\varphi^\mathrm{L}_\mathrm{v}(\bm{j},\bm{r}^\mathrm{L}_{0}),\\
    F_\mathrm{s1}(\bm{r}^\mathrm{L}_{0})&=\frac{1}{N^\mathrm{L}_{x}N_{y}}\sum_{j_x,j_y}\tanh\left[\frac{|\bm{j}a-\bm{r}^\mathrm{L}_{0}|}{z^\mathrm{L}_{0}}\right]\sin\tilde\varphi^\mathrm{L}_\mathrm{v}(\bm{j},\bm{r}^\mathrm{L}_{0}).
\end{align}%
For completeness, we also note that the second harmonic Josephson current can be evaluated by considering the second-order Cooper pair tunneling processes:
\begin{align}
    \tilde{I}_{2}(\varphi;\bm{r}^\mathrm{L}_{0})
    &\propto\frac{1}{N^\mathrm{L}_{x}N_{y}}\sum_{j_x,j_y} \mathrm{Im}[\Delta_\mathrm{L}(\bm{j},\bm{r}^\mathrm{L}_{0})\Delta^{*}_\mathrm{R}\Delta_\mathrm{L}(\bm{j},\bm{r}^\mathrm{L}_{0})\Delta^{*}_\mathrm{R}]\\
    &=\sin 2\varphi
    \frac{|\Delta_{0}|^{4}}{N^\mathrm{L}_{x}N_{y}}\sum_{j_x,j_y}\tanh^2\left[\frac{|\bm{j}a-\bm{r}^\mathrm{L}_{0}|}{z^\mathrm{L}_{0}}\right]
    \cos 2\tilde\varphi^\mathrm{L}_\mathrm{v}(\bm{j},\bm{r}^\mathrm{L}_{0})\notag\\
    &+\cos 2\varphi\frac{|\Delta_{0}|^{4}}{N^\mathrm{L}_{x}N_{y}}\sum_{j_x,j_y}\tanh^2\left[\frac{|\bm{j}a-\bm{r}^\mathrm{L}_{0}|}{z^\mathrm{L}_{0}}\right]\sin 2\tilde\varphi^\mathrm{L}_\mathrm{v}(\bm{j},\bm{r}^\mathrm{L}_{0}).\notag
\end{align}%
Then, we can also find
\begin{align}
    F_\mathrm{c2}(\bm{r}^\mathrm{L}_{0})&=\frac{1}{N^\mathrm{L}_{x}N_{y}}\sum_{j_x,j_y}\tanh^{2}\left[\frac{|\bm{j}a-\bm{r}^\mathrm{L}_{0}|}{z^\mathrm{L}_{0}}\right]\cos 2\tilde\varphi^\mathrm{L}_\mathrm{v}(\bm{j},\bm{r}^\mathrm{L}_{0}),\\
    F_\mathrm{s2}(\bm{r}^\mathrm{L}_{0})&=\frac{1}{N^\mathrm{L}_{x}N_{y}}\sum_{j_x,j_y}\tanh^{2}\left[\frac{|\bm{j}a-\bm{r}^\mathrm{L}_{0}|}{z^\mathrm{L}_{0}}\right]\sin 2\tilde\varphi^\mathrm{L}_\mathrm{v}(\bm{j},\bm{r}^\mathrm{L}_{0}).
\end{align}%
Since both $I_{2}\propto F_\mathrm{c2}$ and $J_{2}\propto F_\mathrm{s2}$ are the $|\Delta_{0}|^{4}$ order, they are generally smaller in amplitude than $I_{1}$ and $J_{1}$, as they are proportional to $|\Delta_{0}|^{2}$.
Then, close to the nodal lines of the first harmonics, the second harmonic components are comparable in amplitude.